\documentclass[%
 reprint,
 nofootinbib,
 amsmath,amssymb,
 aps,
 prd,
]{revtex4-2}
\usepackage[hidelinks]{hyperref}
\usepackage{graphicx}
\usepackage{tabularx}
\begin{document}

\title{Adiabatic gravitational waveform model for compact objects undergoing quasi-circular inspirals into rotating massive black holes}

\author{Zachary Nasipak}
\affiliation{NASA Goddard Space Flight Center, 8800 Greenbelt Road, Greenbelt, Maryland, 20771, USA}
\email{zachary.nasipak@nasa.gov}

\date{\today}

\begin{abstract}
We present \texttt{bhpwave}: a new Python-based, open-source tool for generating the gravitational waveforms of stellar-mass compact objects undergoing quasi-circular inspirals into rotating massive black holes. These binaries, known as extreme-mass-ratio inspirals (EMRIs), are exciting mHz gravitational wave sources for future space-based detectors such as LISA. Relativistic models of EMRI gravitational wave signals are necessary to unlock the full scientific potential of mHz detectors, yet few open-source EMRI waveform models exist. Thus we built \texttt{bhpwave}, which uses the adiabatic approximation from black hole perturbation theory to rapidly construct gravitational waveforms based on the leading-order inspiral dynamics of the binary. In this work, we present the theoretical and numerical foundations underpinning \texttt{bhpwave}. We also demonstrate how \texttt{bhpwave} can be used to assess the impact of EMRI modeling errors on LISA gravitational wave data analysis. In particular, we find that for retrograde orbits and slowly-spinning black holes we can mismodel the gravitational wave phasing by as much as $\sim 10$ radians without significantly biasing EMRI parameter estimation.
\end{abstract}

\maketitle

\section{Introduction}
\label{sec:intro}

Extreme-mass-ratio inspirals (EMRIs) are binaries composed of a compact object with mass $\mu \sim 10 M_\odot$ inspiraling into a massive black hole with mass $M \sim 10^6 M_\odot$. They emit gravitational waves in the milliHertz (mHz) band for months to years, making them promising sources for future space-based detectors, such as the Laser Interferometer Space Antenna (LISA) \cite{NASA11, ESA12}. The prolonged evolution and rich harmonic structure of an EMRI waveform communicates a wealth of information about the binary. Thus, we expect LISA to measure the masses, spins, and orbital characteristics of observed EMRIs with unprecedented accuracy \cite{BabaETC17}. These measurements will provide novel observations of massive black holes and their surrounding environments, while also facilitating high-precision tests of general relativity \cite{BerrETC19}. Extracting this information from an observed signal, however, will likely require subradian-accurate models of EMRI gravitational wave emission.

Due to their small mass ratios $\epsilon = \mu/M \ll 1$, EMRIs are naturally modeled by black hole perturbation theory and the self-force formalism \cite{PounWard20}. Within this framework, the small body is treated as a perturbation to a background Kerr spacetime. The inspiral of the small body is then driven by a \emph{gravitational self-force} (GSF), which arises from the small body interacting with its own perturbation of the spacetime metric. The metric perturbations and the GSF are constructed perturbatively, order by order in $\epsilon$, to derive the dynamics and resulting gravitational waves radiated by the binary.

We can understand the relative impact of these self-forces on EMRI gravitational waveforms by expanding the phasing of the gravitational wave signal $\Phi_\mathrm{GW}$ in powers of $\epsilon$ via a two-timescale analysis \cite{HindFlan08},
\begin{align} \label{eqn:twotimescale}
\Phi_\mathrm{GW} = \frac{1}{\epsilon} \left[ \Phi_\mathrm{0PA} + \epsilon^{1/2} \Phi_\mathrm{res} + \epsilon \Phi_\mathrm{1PA} + O(\epsilon^2)\right].
\end{align}
The leading-order \emph{adiabatic} term\footnote{This term is also referred to as the zeroth post-adiabatic order, post-0 adiabatic, or 0PA term to remain consistent with the naming conventions of the sub-leading terms (e.g., $\Phi_\mathrm{1PA}$).} $\Phi_\mathrm{0PA}$ only depends on the time-averaged \emph{dissipative} first-order GSF, which drives the dissipation of energy and angular momentum from the system. The sub-leading first post-adiabatic order (1PA) piece $\Phi_\mathrm{1PA}$ depends on the remaining contributions from the first-order GSF, corrections due to the spin of the smaller body, and the time-averaged components of the second-order GSF. The half-order correction $\Phi_\mathrm{res}$ arises due to the presence of self-forced $r\theta$-resonances experienced by EMRIs undergoing eccentric and inclined inspirals \cite{FlanHind12}. EMRI gravitational models must, therefore, include the $\Phi_\mathrm{0PA}$, $\Phi_\mathrm{res}$, and $\Phi_\mathrm{1PA}$ effects in order to maintain sub-radian phase accuracy and enable the full scientific potential of future mHz gravitational wave observatories.

Nonetheless purely adiabatic models---which only include $\Phi_\mathrm{0PA}$ in the gravitational wave phasing---may be suitable for detecting (and possibly characterizing) EMRI signals \cite{GairJone06} and are invaluable tools for developing data analysis pipelines for EMRI search and parametrization. They capture almost all of the phasing information and relativistic behavior of EMRI signals, making them powerful probes of astrophysical parameter space. The theoretical foundation and numerical calculation of adiabatic waveforms is also well understood, and has been performed for eccentric, precessing\footnote{In a frame where the massive black hole's angular momentum is held fixed, precession is instead described in terms of the inclination of the small body relative to the equatorial plane of the more massive black hole.} EMRIs \cite{HughETC21}.

However, in practice it is challenging to optimize adiabatic EMRI waveform calculations in order to make them efficient and accessible, so that they can be incorporated in large scale samplings of parameter space. Currently, there is one main open-source self-force model for producing adiabatic EMRI waveforms: the Fast EMRI Waveforms (FEW) Python package \cite{ChuaGallVall19, ChuaETC20, KatzETC20, KatzETC21}. While this tool is the gold standard for EMRI waveform models and data analysis, particularly due to its ability to leverage GPUs, it is currently restricted to eccentric binaries with non-rotating black holes. Since we expect almost all astrophysical EMRIs to possess a rotating massive black hole, it is important to extend models such as FEW into this crucial area of parameter space.

Therefore we introduce \texttt{bhpwave}\footnote{\href{https://bhpwave.readthedocs.io}{https://bhpwave.readthedocs.io}} an open-source Python-based waveform generator that models the adiabatic dynamics and gravitational wave signals produced by a small body undergoing a quasi-circular\footnote{By ``quasi-circular," we mean that, at any moment of time, the inspiral motion is tangent to a circular geodesic as described in Sec.~\ref{sec:adiabatic}. This is a consequence of the fact that in the adiabatic limit, circular orbits remain circular, i.e., $e(t) = O(\epsilon)$ \cite{KennOri96, Kenn98}.} (non-eccentric, non-precessing) inspiral into a {rotating} massive black hole. This code can model any binary with an initial orbital separation of $r_0 \leq 60 M$ and a massive black hole spin in the range $|a| \leq 0.9999 M$. (See Sec.~\ref{sec:adiabatic} for exact definitions of $a$ and $r_0$.) Furthermore, the model supports any mass-ratio, though adiabatic waveform models are most relevant for $\epsilon \lesssim 10^{-4}$. By leveraging parallel computations across CPUs, \texttt{bhpwave} can evaluate years-long waveforms in milliseconds. Even on a standard laptop, waveform evaluations still complete within a few hundred milliseconds to a couple seconds.

Importantly, \texttt{bhpwave} does not provide the first adiabatic calculation of quasi-circular inspirals. These systems have long been studied in the literature \cite{Detw78, KennOri96, FinnThor00, Hugh00b, GralHughWarb16}, and other works have even modeled eccentric, equatorial and eccentric, inclined (precessing) EMRIs for a variety of massive black hole spins \cite{FujiShib20, SkouLuke22, HughETC21}. Recent works have also produced waveforms that include the effects due to the small body's spin (which is a post-adiabatic effect) \cite{SkouLuke22, DrumETC23}, while effective-one-body models have combined post-Newtonian and GSF information to provide EMRI waveforms for binaries with aligned spins \cite{AlbeETC23}.

Despite this rich literature, it remains a practical challenge to implement an adiabatic model that is efficient, accurate, and accessible across a majority of the parameter space, especially for arbitrary values of the massive black hole spin. Even when neglecting eccentricity, the inclusion of spin presents a number of issues that are less pronounced for binaries with non-spinning bodies, such as the larger range of radial separations accessible to spinning binaries, the faster frequency evolution for more deeply bound orbits, and the higher harmonic content needed for modeling rapidly-rotating systems. By building upon the theoretical foundations described in the literature, our aim for \texttt{bhpwave} is to provide an efficient and accessible model that fills in a new area of parameter space and provides an important stepping stone for developing more advanced waveform codes that incorporate all possible orbital effects.

To encourage the continued development of open-source adiabatic EMRI models, this paper outlines the theoretical and numerical foundations underpinning \texttt{bhpwave}. In Sec.~\ref{sec:adiabatic} we summarize the quasi-circular limit of the adiabatic approximation within the context of black hole perturbation theory to establish methods and notation.
In Sec.~\ref{sec:numerical} we describe both the numerical routines used to generate the adiabatic data for \texttt{bhpwave} and the algorithms employed by \texttt{bhpwave} to generate quasi-circular inspirals, waveform harmonics, and gravitational signals. Additionally, we provide validation tests and model comparisons to verify the accuracy of \texttt{bhpwave}. To demonstrate the utility of \texttt{bhpwave}, in Sec.~\ref{sec:errors} we present example problems for testing the impact of modeling errors on parameter biases for LISA data analysis. Finally, we discuss further applications and possible extensions of \texttt{bhpwave} in Sec.~\ref{sec:conclusion}. To support community collaboration and open-science, the codes for calculating all of the data, performing all of the analyses, and generating all of the plots presented in this work are made available via the public Github repository \texttt{bhpwave-article}.\footnote{\href{https://github.com/znasipak/bhpwave-article}{https://github.com/znasipak/bhpwave-article}} For this paper we use the metric signature $(-+++)$, the sign conventions, where applicable, of \cite{MisnThorWhee73}, and units such that $G=c=1$.

\section{Adiabatic approximation}
\label{sec:adiabatic}

We provide a brief summary of the leading-order adiabatic approximation of a small body's quasi-circular inspiral into a rotating massive black hole. (For extensive discussions of the point-particle and adiabatic approximations see \cite{Hugh00b, Mino03, DrasHugh06, HughETC21} and references therein.) At zeroth-order the small body follows a circular, equatorial geodesic in Kerr spacetime (see \ref{sec:geo}). Due to its mass and motion, the small body excites gravitational radiation. At leading-order this radiation is captured by non-zero perturbations to the Weyl scalar $\psi_4$, which we construct via the Teukolsky equation (see \ref{sec:fluxes}). The resulting flux of energy dissipated via gravitational wave emission---which we compute from $\psi_4$---then drives the decay of the small body's orbital energy and thus the adiabatic inspiral (see \ref{sec:inspiral}). From this inspiral, we then construct the adiabatic gravitational waveform (see \ref{sec:waveform}).

\subsection{Circular, equatorial geodesics}
\label{sec:geo}

Consider a small body with mass $\mu$ orbiting in a Kerr spacetime with metric $g_{\mu\nu}$, which, in Boyer-Lindquist coordinates $(t,r,\theta,\phi)$, is defined by the line element
\begin{multline}
    ds^2 = -\left(1 - \frac{2Mr}{\Sigma} \right) dt^2 - \frac{4Ma r \sin^2\theta}{\Sigma} dtd\phi + \frac{\Sigma}{\Delta} dr^2
    \\
    + \Sigma d\theta^2 + {\sin^2\theta}\left(r^2+a^2 + \frac{2Ma^2r\sin^2\theta}{\Sigma} \right) d\phi^2,
\end{multline}
where $a$ is the Kerr spin parameter, $M$ is the Kerr mass parameter, $\Delta = r^2 - 2Mr + a^2$, and $\Sigma = r^2+a^2\cos^2\theta$.

The mass $\mu$ follows a geodesic $z_p^\mu \doteq (t_p, r_p, \theta_p, \phi_p)$ that maintains a constant Boyer-Lindquist radius $r_p = r_0$ and is restricted to the equatorial plane $\theta_p = \pi/2$ (with respect to the angular momentum of the Kerr black hole). Due to the Killing symmetries of Kerr spacetime, this motion possesses three constants of motion: the orbital energy $\mathcal{E} = -u_t$, the $z$-component of the orbital angular momentum $\mathcal{L}_z = u_\phi$, and the Carter constant $\mathcal{Q} = Q_{\mu\nu} u^\mu u^\nu$ (see Eq.~(8) in \cite{SteiWarb20} for an explicit definition of $Q_{\mu\nu}$), which are related to $a$ and $r_0$ by
\begin{subequations} \label{eqn:En}
\begin{align}
    \mathcal{E} &= \frac{1 - 2 v^2 \pm \hat{a} v^3}{\sqrt{1-3v^2\pm 2\hat{a}v^3}},
    \\
    \mathcal{L}_z &= \pm M  \frac{1 \mp 2 \hat{a} v^3 + \hat{a}^2 v^4}{v\sqrt{1-3v^2\pm 2\hat{a}v^3}},
    \\
    \mathcal{Q} &= 0,
\end{align}
\end{subequations}
where $\hat{a} \equiv a/M$, $v^2 = M/r_0$, and $+$ ($-$) refers to prograde (retrograde) orbits.

For circular orbits, the four-velocity $u^\alpha = dz_p^\alpha/d\tau \doteq (\omega_t, 0, 0, \omega_\phi)$ is constant along the geodesic, where $\tau$ is the proper time of the small body. The rates at which (Boyer-Lindquist) time and the azimuthal angle accumulate with $\tau$ are given by
\begin{subequations}
\begin{align}
    \omega_t &= \frac{g_{\phi\phi} \mathcal{E} + g_{t\phi}\mathcal{L}_z}{g_{t\phi}^2-g_{\phi\phi}g_{tt}} = \frac{1 \pm \hat{a} v^3}{\sqrt{1 - 3 v^2 \pm 2 \hat{a} v^3}},
    \\
    \omega_\phi &= -\frac{g_{t\phi} \mathcal{E} + g_{tt}\mathcal{L}_z}{g_{t\phi}^2-g_{\phi\phi}g_{tt}} = \pm \frac{v^3}{M\sqrt{1 - 3 v^2 \pm 2 \hat{a} v^3}},
\end{align}
\end{subequations}
respectively. Straightforward integration then yields
\begin{align*}
    t_p(\tau) &= \omega_t \tau,
    &
    r_p(\tau) &= r_0,
    &
    \theta_p(\tau) &= \frac{\pi}{2},
    &
    \phi_p(\tau) &= \omega_\phi\tau.
\end{align*}
Combining these results, we can re-express the evolution of the orbital phase in terms of coordinate time, $\phi_p(t) = \Omega_p t$, where the geodesic orbital frequency is given as
\begin{equation} \label{eqn:OmegaOfR}
    \Omega_p = \frac{\omega_\phi}{\omega_t}
    = \pm \frac{v^3}{M(1 \pm \hat{a} v^3)}.
\end{equation}
Finally, the innermost stable circular orbit (ISCO) is defined in terms of the minimum radius
\begin{align}
    r_\mathrm{ISCO} &= 3 + z_2 \mp
    \sqrt{(3 - z_1)(3 + z_1 + 2z_2)},
    \\ \notag
    z_1 & = 1 + (1 - \hat{a}^2)^{1/3}
    \left((1-\hat{a})^{1/3} + (1 + \hat{a})^{1/3} \right),
    \\ \notag
    z_2 &= \sqrt{3 \hat{a}^2 + z_1^2},
\end{align}
and maximum frequency $M\Omega_\mathrm{ISCO} = r_\mathrm{ISCO}^{3/2}(M^{3/2} + \hat{a} r_\mathrm{ISCO}^{3/2})^{-1}$.

\subsection{Gravitational wave fluxes}
\label{sec:fluxes}

Next we consider how the small body's motion excites perturbations to the background spacetime. Using the Teukolsky formalism \cite{Teuk73}, we describe these perturbations in terms of the Weyl scalar $\psi_4$, which vanishes in an unperturbed Kerr spacetime and captures two of the ten gravitational degrees of freedom of the metric perturbation. Near infinity, these two degrees of freedom can be related to $h_+$ and $h_\times$ \cite{DrasHugh06}, the two polarizations of the gravitational strain, via
\begin{align} \label{eqn:psi4ToH}
    \psi_4 (r\rightarrow \infty) \simeq \frac{1}{2}(\ddot{h}_+ - i \ddot{h}_\times).
\end{align}
At adiabatic order, $\psi_4$ satisfies the spin-weight $s=-2$ Teukolsky equation (Eq.~(4.7) in \cite{Teuk73}) with a point-particle source (see Sec.~II of \cite{SasaTago03}). The solution is amenable to separation of variables in the frequency-domain, leading to the mode-sum representation
\begin{align} \label{eqn:psi4Modes}
    \psi_4 = \rho^{4} \sum_{j=2}^\infty \sum_{m=-j}^j R_{-2jm}(r) S_{-2jm}(\theta;\gamma) e^{im\phi} e^{-im\Omega_pt},
\end{align}
where $\rho = -(r-ia\cos\theta)^{-1}$, $S_{-2jm}(\theta;\gamma)$ is the spin-weighted spheroidal harmonic of spin-weight $s=-2$ and spheroidicity $\gamma=a m\Omega_p$ (which satisfies Eq.~(2.7) in \cite{TeukPres74}), and $R_{-2jm}(r)$ is the $s=-2$ radial Teukolsky solution (which satisfies Eq.~(3.12) in \cite{DrasHugh06}).

The spin-weighted spheroidal harmonics can be conveniently represented as a rapidly-converging series of spin-weighted spherical harmonics $Y_{s\ell m}(\theta)e^{im\phi}$ \cite{Hugh00b},
\begin{align}
    S_{sjm}(\theta; \gamma) = \sum_{\ell=\ell_\mathrm{min}}^\infty b_{sjm}^\ell(\gamma) Y_{s\ell m}(\theta),
\end{align}
where $\ell_\mathrm{min}=\mathrm{max}[|m|,|s|]$. This decomposition is particularly useful because we can then reproject $\psi_4$ onto an angular basis that is independent of frequency,
\begin{align} \label{eqn:psi4LModes}
    \psi_4 = \rho^{4} \sum_{\ell=2}^\infty \sum_{m=-\ell}^\ell X_{-2\ell m}(r) Y_{-2\ell m}(\theta)e^{im(\phi - \Omega_pt)},
\end{align}
where
\begin{align} \label{eqn:XfromR}
    X_{-2\ell m}(r) = \sum_{j=\ell_\mathrm{min}}^\infty b_{-2jm}^\ell(\gamma)R_{-2jm}(r).
\end{align}
Furthermore, the spin-weighted spherical harmonics form a complete and orthonormal set of basis functions on the unit sphere (for the same value of spin-weight),
\begin{align}
    \int Y_{s\ell m} {Y}_{s\ell'm'} e^{i(m-m')\phi} d\Omega = \delta_{l l'} \delta_{m m'}.
\end{align}



For our point-particle source on a circular geodesic, the radial mode function $R_{-2jm}(r)$ have the asymptotic form
\begin{subequations} \label{eqn:asympR}
\begin{align}
    R_{-2jm}(r\rightarrow r_+) &\simeq Z^\mathcal{H}_{-2jm} \Delta^{2} e^{-im k_p r_*},
    \\
    R_{-2jm}(r\rightarrow \infty) &\simeq Z^\mathcal{I}_{-2jm} r^{3} e^{+im\Omega_p r_*},
\end{align}
\end{subequations}
where $k_p = \Omega_p - \Omega_+$, the horizon frequency $\Omega_+ = a/(2Mr_+)$, and the tortoise coordinate $r_*$ is given by the differential relation $dr_*/dr = (r^2+a^2)/\Delta$.
The amplitudes $Z^\mathcal{H/I}_{-2jm}$ are often referred to as \emph{Teukolsky amplitudes} and are constructed via the standard Green's function method, also known as the method of variation of parameters (see Sec.~III\,A in \cite{HughETC21} or \cite{DrasHugh06} for further details). From \eqref{eqn:XfromR}, we see that $X_{-2\ell m}$ then possesses the same asymptotic behavior as $R_{-2\ell m}$ in \eqref{eqn:asympR} but with modified amplitudes
\begin{align}
    Z^\mathcal{H/I}_{-2jm} \rightarrow X^\mathcal{H/I}_{-2\ell m} = \sum_{j=\ell_\mathrm{min}}^\infty b^\ell_{-2jm}(\gamma) Z^\mathcal{H/I}_{-2jm}.
\end{align}
As a final note, because $Z^\mathcal{H/I}_{-2jm}$ and  $X^\mathcal{H/I}_{-2\ell m}$ depend on the source motion, we can parametrize both amplitudes in terms of the orbital constants, e.g., $X^\mathcal{H/I}_{-2\ell m}= {X}^\mathcal{H/I}_{-2\ell m}(\Omega_p; a)$.

Upon obtaining $\psi_4$, we can then calculate the flux of energy that $\psi_4$ radiates away to infinity $\dot{E}^\mathcal{I}$ and down the black hole horizon $\dot{E}^\mathcal{H}$ \cite{TeukPres74},
\begin{subequations} \label{eqn:fluxes}
\begin{align}
    \dot{E}^\mathcal{I} &= \frac{1}{4\pi} \sum_{jm} \alpha_{jm}^\infty {|Z^\mathcal{I}_{-2jm}|^2},
    \\
    \dot{E}^\mathcal{H} &= \frac{1}{4\pi} \sum_{jm} \alpha_{jm}^\mathcal{H} {|Z^\mathcal{H}_{-2jm}|^2},
\end{align}
\end{subequations}
where
\begin{align*}
    \alpha_{jm}^\mathcal{I} &= \frac{1}{m^2\Omega_p^2},
    \\
    \alpha_{jm}^\mathcal{H} &= \frac{256m^2 k_p\Omega_p(2Mr_+)^5(m^2k_p^2+4\epsilon^2)(m^2k_p^2+16\epsilon^2)}{|C_{2jm}|^2},
\end{align*}
$\epsilon = (r_+ - M)/(2Mr_+)$, and the Teukolsky-Starobinsky constant is given by
\begin{align}
    |C_{2jm}|^2 &=\Lambda_{2j m }^2(\Lambda_{2j m }-2)^2
    \\ \notag
    & \quad + 8 a m^2 \Omega_p(1 - a \Omega_p)(\Lambda_{2j m }-2)(5\Lambda_{2j m }-4)
    \\ \notag
    & \quad \quad
    +48 (a m \Omega_p)^2[4(\Lambda_{2j m }-1)+3m(1- a \Omega_p)].
\end{align}
Note that the Chandrasekhar eigenvalue $\Lambda_{sjm} = \lambda_{sjm} + s(s+1)$ is invariant under the interchange $s\rightarrow -s$ \cite{Chan83}.
The angular momentum fluxes are then related to the energy fluxes by $\dot{L}_z^{\mathcal{I}/\mathcal{H}} = \Omega^{-1} \dot{E}^{\mathcal{I}/\mathcal{H}}$, leading to the total gravitational wave fluxes
\begin{align}
    \dot{E}^\mathrm{GW} &= \dot{E}^\mathcal{I} + \dot{E}^\mathcal{H},
    &
    \dot{L}_z^\mathrm{GW} &= \dot{L}_z^\mathcal{I} + \dot{L}_z^\mathcal{H}.
\end{align}

\subsection{Adiabatic quasi-circular inspirals}
\label{sec:inspiral}

Due to gravitational {radiation-reaction}, the small body does not remain on a circular geodesic. The binary {radiates} gravitational waves and the small body {reacts} to the radiative losses of energy and angular momentum by undergoing a quasi-circular inspiral into the rotating massive black hole. We can parametrize this motion in terms of the time-evolving orbital energy ${E}(t)$ and orbital phase $\Phi(t)$ of the inspiraling body, which at leading order is given by the equations of motion \cite{PounWard20,HughETC21},
\begin{align} \label{eqn:eom}
    \frac{d {E}}{dt} &= -\epsilon \mathcal{F}_E + O(\epsilon^2),
    &
    \frac{d\Phi}{dt} &= \Omega + O(\epsilon).
\end{align}
Because $\dot{E} \sim \epsilon$, the orbital energy evolves gradually and the evolution can be understood in terms of a time-averaged \emph{osculating geodesics} method \cite{GairETC11}. At any time $t_0$, the motion is approximately tangent to a geodesic with energy $E(t_0) \approx \mathcal{E}$ and frequency $\Omega(t_0) \approx \Omega_p$. The small body then evolves from geodesic to geodesic based on its gravitational wave emission until the body approaches the ISCO, at which point the approximation breaks down and we end the evolution.

In effect, at leading order we can express the orbital frequency $\Omega$ in terms of the orbital energy ${E}$ using Eqs.~\eqref{eqn:En} and \eqref{eqn:OmegaOfR}, but with $\mathcal{E}$ and $\Omega_p$ replaced by $E$ and $\Omega$, respectively. Furthermore, the forcing term $\mathcal{F}_E$ is constructed using flux-balance arguments \cite{Mino03, Galt82, QuinWald99}: for a point-particle on a circular geodesic the loss of orbital energy is balanced by the gravitational wave energy flux $\dot{E}^\mathrm{GW}$, leading to the relation $\epsilon \mathcal{F}_E = \dot{E}^\mathrm{GW}$.

As seen from Eq.~\eqref{eqn:eom}, the time it takes the system to undergo this inspiral scales with the radiation reaction timescale $T_\mathrm{insp} \sim T_\mathrm{rr} = M \epsilon^{-1}$. Similarly, the total accumulated phase scales like $\Delta\Phi_\mathrm{insp} \sim \epsilon^{-1}$, thus giving the leading-order phase contribution to the gravitational wave signal, as described in Eq.~\eqref{eqn:twotimescale}.

\subsection{Time domain adiabatic waveform}
\label{sec:waveform}

After determining the adiabatic inspiral of the smaller body, we generalize \eqref{eqn:psi4LModes}---our geodesic expression for $\psi_4$---by once again leveraging the fact that at any moment of time the motion is approximately tangent to a geodesic. Consequently, the field amplitudes are promoted from constants to quantities that slowly evolve with frequency and time $X^\mathcal{H/I}_{-2\ell m}(\Omega_p; a) \rightarrow X^\mathcal{H/I}_{-2\ell m}(\Omega(t); a)$, while the field's phase rapidly accumulates in proportion to the orbital phase $\Omega_p t \rightarrow \Phi(t)$. As a result, the adiabatic waveform $h = h_+ - i h_\times$ takes the form
\begin{subequations} \label{eqn:adiabaticWaveform}
\begin{align}
    h(u,r,\theta,\phi) &= \frac{\mu}{r}\sum_{\ell m} h_{\ell m}(u,r,\theta,\phi),
    \\
    h_{\ell m}(u,r,\theta,\phi) & = B_{\ell m}(u) Y_{-2\ell m}(\theta) e^{im[\phi - \Phi(u)]}
\end{align}
\end{subequations}
where, assuming $r \gg M$, the waveform amplitudes are given by
\begin{align} \label{eqn:amplitude}
    B_{\ell m}(u) = -2\frac{X_{-2\ell m}^\mathcal{I}( \Omega(u);a)}{m^2\Omega^2(u)} = A_{\ell m} e^{i \psi_{\ell m}}.
\end{align}
For later convenience we introduce the magnitude of the waveform amplitudes $A_{\ell m} = |B_{\ell m}|$ and the phase of the complex amplitudes $\psi_{\ell m}$, both of which depend on $\Omega(u)$ and $a$. Furthermore, rather than using the conventions of \cite{KatzETC21,HughETC21}, we follow \cite{PounWard20} and parametrize the adiabatic waveform in terms of the outgoing time-coordinate $u=t-r_*$. As a result, our expression is consistent with the adiabatic expression obtained from a full two-timescale analysis, and if we hold the orbital constants and frequencies fixed, \eqref{eqn:adiabaticWaveform} reduces to a geodesic ``snapshot'' waveform obtained via \eqref{eqn:psi4ToH} and \eqref{eqn:psi4LModes}. In practice, when implementing \eqref{eqn:adiabaticWaveform} in \texttt{bhpwave}, we replace $u$ with Boyer-Lindquist time $t$ to match the FEW model (which is described by (10) in \cite{KatzETC21}). This is equivalent to parametrizing all of our waveforms with an initial time $t_0 = r_*$. Therefore, we replace $u$ with $t$ for the remainder of this work.

\subsection{Frequency domain waveforms}

The amplitude and phase decomposition of \eqref{eqn:adiabaticWaveform} makes it particularly straightforward to represent our waveforms in the frequency domain via the stationary phase approximation \cite{HughETC21},
\begin{subequations} \label{eqn:fourierH}
\begin{align}
    \tilde{h}(f) &= \int_{-\infty}^\infty h(t) e^{2\pi i f t} dt =  \frac{\mu}{r}\sum_{\ell m}\tilde{h}_{\ell m}(f),
    \\
    \tilde{h}_{\ell m}(f) & \approx \tilde{B}_{\ell m}(f) Y_{-2\ell m} e^{i[2\pi f t_p(f)+ m(\phi-\Phi[t_p(f)])]},
\end{align}
\end{subequations}
where
\begin{align} \label{eqn:fourierAmplitude}
    \tilde{B}_{\ell m}(f) \approx \sqrt{\frac{2\pi}{i m \dot{\Omega}[t_p(f)]}} {B}_{\ell m}(t_p(f)),
\end{align}
and $t_p(f)$ refers to the times at which the binary emits gravitational waves with frequency $f$. For each $(\ell, m)$-mode, time and frequency are (approximately) related by
\begin{align} \label{eqn:fourierPhase}
    2\pi f \approx m \dot{\Phi}(t) = m \Omega(t),
\end{align}
which we can invert to obtain $t_p(f)$ for individual harmonics. Both \eqref{eqn:fourierAmplitude} and \eqref{eqn:fourierPhase} neglect terms related to $\dot{\psi}_{\ell m}$ and $\ddot{\psi}_{\ell m}$, which introduces an $O(\epsilon)$ error to the phase and an $O(\epsilon^{1/2})$ error to the amplitude. (See Appendix \ref{app:fourierPhase} for further details.) Since the amplitudes scale as $\tilde{B}_{\ell m}(f) \sim 1/\sqrt{\epsilon}$, it is safe to neglect these terms for small mass-ratios.

\subsection{Solar system barycenter waveforms}

Up to this point, waveforms have been constructed in the source frame, with our Boyer-Lindquist coordinate system centered on the massive black hole. To get the observed waveform in the solar system barycenter (SSB) frame $h_\mathrm{SSB}$, we adopt the same conventions as the \textsc{FEW} model and make use of the transformation provided in \cite{KatzETC21}. In this frame,
a generic EMRI system is parametrized by 12 intrinsic parameters $(M, \mu, a, \vec{a}_2, p_0, e_0, x_0, \Phi_{r0}, \Phi_{\theta 0}, \Phi_{\phi 0})$---where $\vec{a}_2$ is the spin-vector of the smaller body; $p_0$, $e_0$, and $x_0$ are respectively the initial semi-latus rectum, orbital eccentricity, and projection of the orbital inclination; and $\Phi_{r0}$, $\Phi_{\theta 0}$, $\Phi_{\phi 0}$ are respectively the initial radial, polar, and azimuthal phases---and by 5 extrinsic parameters $(d_L, q_S, \phi_S, q_K, \phi_K)$---where $d_L$ is the luminosity distance to the source EMRI, $q_S$ and $\phi_S$ are the polar and azimuthal sky positions of the source, and $q_K$ and $\phi_K$ are the polar and azimuthal angles defining the orientation of the massive black hole's axis of rotation. In our simplified quasi-circular setup, five of the intrinsic parameters are constrained to the values $\vec{a}_2 = \vec{0}$, $e_0 = 0$, $|x_0| = 1$, $\Phi_{r0} = \Phi_{\theta 0} = 0$, while two of the remaining free intrinsic parameters are given by $\Phi_{\phi 0} = \Phi(0)$ and $p_0 = r_0$.

The extrinsic parameters are related to Boyer-Lindquist $(r,\theta, \phi)$ via
\begin{subequations} \label{eqn:SSBtoSource}
    \begin{align}
        r &= d_L,
        \\
        \cos\theta &= - \cos q_S\cos q_K
        \\ \notag
        & \qquad \quad -\sin q_S \sin q_K \cos(\phi_K - \phi_S),
        \\
        \phi &= -\frac{\pi}{2}.
    \end{align}
\end{subequations}
The frame transformation also rotates the polarization basis by the polarization angle $\psi$ \cite{KatzETC21}. As a result, $h_\mathrm{SSB}$ is related to $h$, the strain in the source frame, via
\begin{align} \label{eqn:polarization}
    h_\mathrm{SSB} = e^{2i\psi} h = \frac{D_\psi^2}{|D_\psi|^2} h = \frac{D_\psi}{D^*_\psi} h,
\end{align}
where $D_\psi$ is given by
\begin{multline}
    D_\psi = \cos q_S \sin q_K \cos(\phi_K - \phi_S)
    \\
     - \sin q_S\cos q_K +i \sin q_K \sin(\phi_K - \phi_S),
\end{multline}
and $a^*$ denotes the complex conjugate of $a$. Note that there are two cases in which the preferred wave basis in \cite{KatzETC21} is no longer uniquely defined: $(q_K, \phi_K) = (q_S, \phi_S)$ and $(q_K, \phi_K) = (\pi - q_S, \pi + \phi_S)$. In these instances the SSB and source frames are aligned or anti-aligned with one another, leading to $D_\psi = 0$ and invalidating the last equality in \eqref{eqn:polarization}. Therefore, we set $h_\mathrm{SSB} = h$ if $(q_K, \phi_K) = (q_S, \phi_S)$ or $(q_K, \phi_K) = (\pi - q_S, \pi + \phi_S)$.

\section{Numerical methods}
\label{sec:numerical}

In the following section we discuss the numerical implementation of \texttt{bhpwave}, in particular our process for solving the equations of motion (Secs.~\ref{sec:eom} and \ref{sec:traj}), constructing the harmonic amplitudes of the gravitational wave modes (Sec.~\ref{sec:amp}), and evaluating the total waveform signal (Sec.~\ref{sec:wave}). Finally we compare our model, \texttt{bhpwave}, to the FastEMRIWaveform (FEW) model to test its accuracy. (Sec.~\ref{sec:few}).

\subsection{Equations of motion and the numerical domain}
\label{sec:eom}

To simplify the calculation of our inspirals, we introduce the dimensionless and rescaled orbital quantities
\begin{align}
    \check{t} &= \frac{\mu t}{M^2},
    &
    \check{\Phi} &= \frac{\mu \Phi}{M},
    &
    \hat{\Omega} &= M \Omega,
    &
    \hat{a} &= \frac{a}{M}.
\end{align}
A hat represents a quantity that is made dimensionless, while a check represents a quantity that is dimensionless and scaled by the mass ratio. We then reparametrize the equations of motion in terms of the (dimensionless) orbital frequency, yielding
\begin{subequations} \label{eqn:eom1}
\begin{align} \label{eqn:dOmegadT}
    \frac{d\hat{\Omega}}{d\check{t}} &= -\left[\partial_{\mathcal{E}}\hat{\Omega}(\hat{\Omega}; \hat{a}) {\mathcal{F}}_E(\hat{\Omega}; \hat{a})\right],
    \\ \label{eqn:dPhidT}
    \frac{d\check{\Phi}}{d\check{t}} &= \hat{\Omega},
\end{align}
\end{subequations}
for time-domain trajectories, and
\begin{subequations} \label{eqn:eom2}
\begin{align} \label{eqn:dtdOmega}
    \frac{d\check{t}}{d\hat{\Omega}} &= -\left[\partial_{\mathcal{E}}\hat{\Omega}(\hat{\Omega}; \hat{a}) {\mathcal{F}}_E(\hat{\Omega}; \hat{a})\right]^{-1},
    \\ \label{eqn:dPhidOmega}
    \frac{d\check{\Phi}}{d\hat{\Omega}} &= -\hat{\Omega} \left[\partial_{\mathcal{E}}\hat{\Omega}(\hat{\Omega}; \hat{a}) {\mathcal{F}}_E(\hat{\Omega}; \hat{a})\right]^{-1},
\end{align}
\end{subequations}
for the frequency domain.
The Jacobian $\partial_{\mathcal{E}}\hat{\Omega}$ is analytically obtained from \eqref{eqn:En} and \eqref{eqn:OmegaOfR}, while we construct ${\mathcal{F}}_E$ numerically via \eqref{eqn:fluxes}.

In this form, the equations of motion no longer depend on the masses of the binary, just the dimensionless orbital frequency and dimensionless Kerr spin parameter. By parametrizing in terms of frequency in \eqref{eqn:eom2}, we decouple the evolution in time and phase, leading to solutions $\check{t}(\hat{\Omega}; \hat{a})$ and $\check{\Phi}(\hat{\Omega}; \hat{a})$. In this form, it is straightforward to construct the frequency domain waveforms via Eqs.~\eqref{eqn:fourierH}-\eqref{eqn:fourierPhase}.

\begin{figure}
    \centering
    \includegraphics[width=0.95\columnwidth]{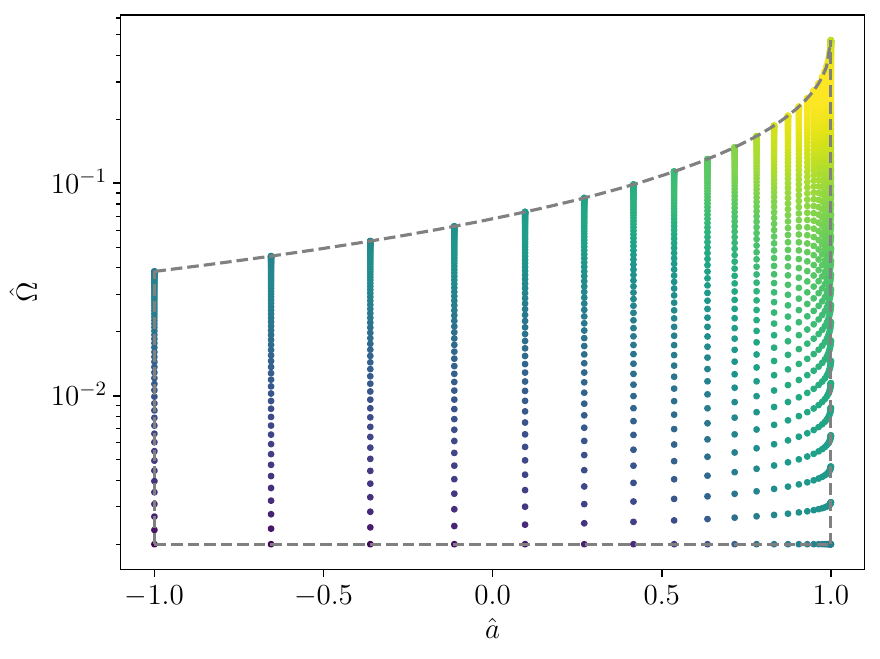}
    \caption{A $32\times 64$ grid in $(\alpha, \chi)$ mapped to the $(\hat{a}, \hat{\Omega})$ domain.  Each dot represents a sampled point in the parameter space. These points are equally-spaced in $\alpha$ and in $\chi$, but cluster around $\hat{\Omega} = \hat{\Omega}_\mathrm{ISCO}$ and $\hat{a} = \hat{a}_\mathrm{max}$. The shading of each point reflects the relative density of points in the $(\hat{a}, \hat{\Omega})$ domain, while the dashed lines represent the domain boundaries. Note that this is the downsampled version of the numerical grid used for interpolating flux and trajectory information within \texttt{bhpwave}.}
    \label{fig:domain}
\end{figure}

Therefore, we first interpolate the data $\mathcal{F}_E(\hat{\Omega}; \hat{a})$, and then solve the equations of motion for the rescaled frequency-domain trajectories $\check{t}(\hat{\Omega}; \hat{a})$ and $\check{\Phi}(\hat{\Omega}; \hat{a})$. We set the initial conditions $\check{t}(\hat{\Omega}_\mathrm{ISCO}) = \check{\Phi}(\hat{\Omega}_\mathrm{ISCO}) = 0$. Thus all time and phase values are $\leq 0$. We then solve for the time-domain trajectories $\hat{\Omega}(\check{t}; a)$ and $\check{\Phi}(\check{t}; a)$. Data can be precomputed and stored on a numerical grid spanning the domains of $(\hat{a}, \hat{\Omega})$ or $(\hat{a}, \check{t})$. From these grids, we construct bicubic spline interpolants to represent the trajectories, which can be rapidly evaluated to get any quasi-circular inspiral within our domain. For this work, we set the boundaries of our grid at $\hat{a} \in [-\hat{a}_\mathrm{max}, +\hat{a}_\mathrm{max}]$ and $\hat{\Omega} \in [\hat{\Omega}_\mathrm{min}, \hat{\Omega}_\mathrm{ISCO}]$ with $\hat{a}_\mathrm{max} = 0.9999$ and $\hat{\Omega}_\mathrm{min} = 2\times 10^{-3}$ (which corresponds to $r_0 \approx 60 M$). This domain is plotted in Fig.~\ref{fig:domain}.

Note that, due to the one-to-one mapping between time and frequency, we could calculate the frequency-domain trajectories and then invert them to get the time-domain trajectories. In practice, however, we create four separate spline representations for $\check{t}(\hat{\Omega}; \hat{a})$, $\check{\Phi}(\hat{\Omega}; \hat{a})$, $\hat{\Omega}(\check{t}; a)$, and $\check{\Phi}(\check{t}; a)$. This is much more computationally efficient then inverting one set of solutions through root-finding methods (e.g., $\hat{\Omega}(\check{t}_0) = \mathrm{Root}[\check{t}(\hat{\Omega}) - \check{t}_0]$) and avoids the accumulation of interpolation errors when taking the composition of two splines (e.g., $\check{\Phi}(\check{t}_0) = \check{\Phi}[\hat{\Omega}(\check{t}_0)])$.

\subsection{Interpolated fluxes and trajectories}
\label{sec:traj}

\begin{figure*}[!htp]
    \centering
    \includegraphics[width=0.95\linewidth]{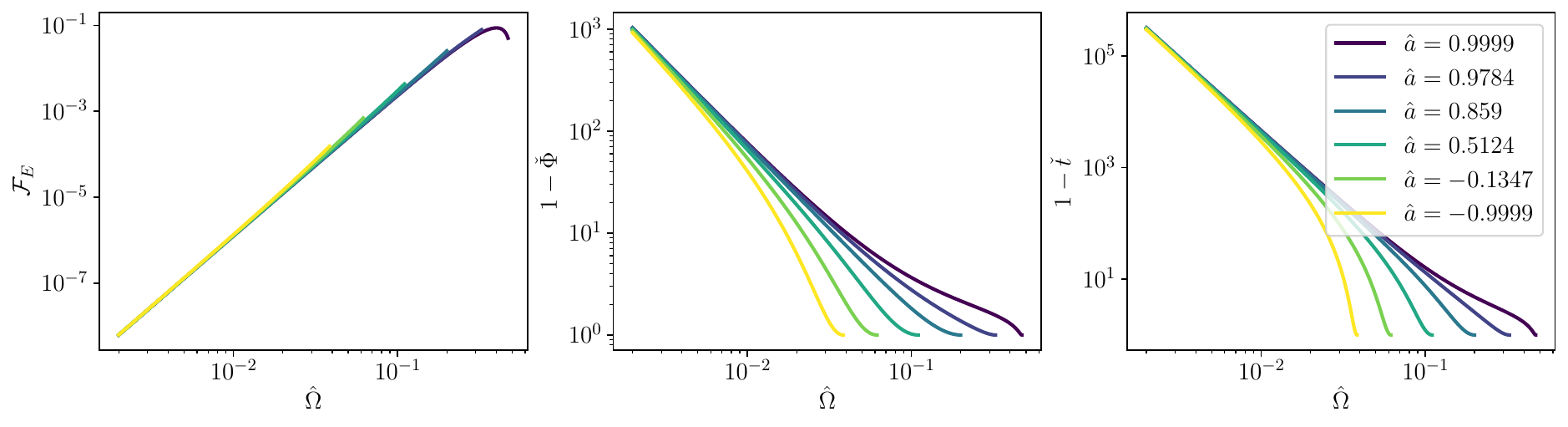}
    \caption{The energy flux $\mathcal{F}_E$ (left panel), (rescaled) orbital phase $\check{\Phi}$ (middle panel), and (rescaled) orbital time to merger $\check{t}$ as a function of (dimensionless) orbital frequency $\hat{\Omega}$. Different colors (shades) correspond with different values of $\hat{a}$. Therefore, lower spins terminate at lower ISCO frequencies. Note that the spin has a very small effect on the flux.}
    \label{fig:original}
\end{figure*}

There are two main challenges to constructing $\mathcal{F}_E(\hat{\Omega}; \hat{a})$ via numerical interpolation: (1) the lower frequency boundary $\hat{\Omega}=\hat{\Omega}_\mathrm{ISCO}$ has a particularly strong dependence on the black hole spin as highlighted in Figure \ref{fig:domain}; and (2) $\mathcal{F}_E$, $\check{t}$, and $\check{\Phi}$ rapidly accumulate as $\hat{\Omega} \rightarrow \hat{\Omega}_\mathrm{ISCO}$ and $\hat{a} \rightarrow 1$ as shown in Figure \ref{fig:original}. A simple uniform sampling in $\hat{a}$ and $\hat{\Omega}-\hat{\Omega}_\mathrm{ISCO}$ could lead to substantial errors in our interpolating functions due to the large magnitudes of the higher-order derivatives with respect to $\hat{a}$ and $\hat{\Omega}$.

\begin{figure*}[!htp]
    \centering
    \includegraphics[width=0.98\linewidth]{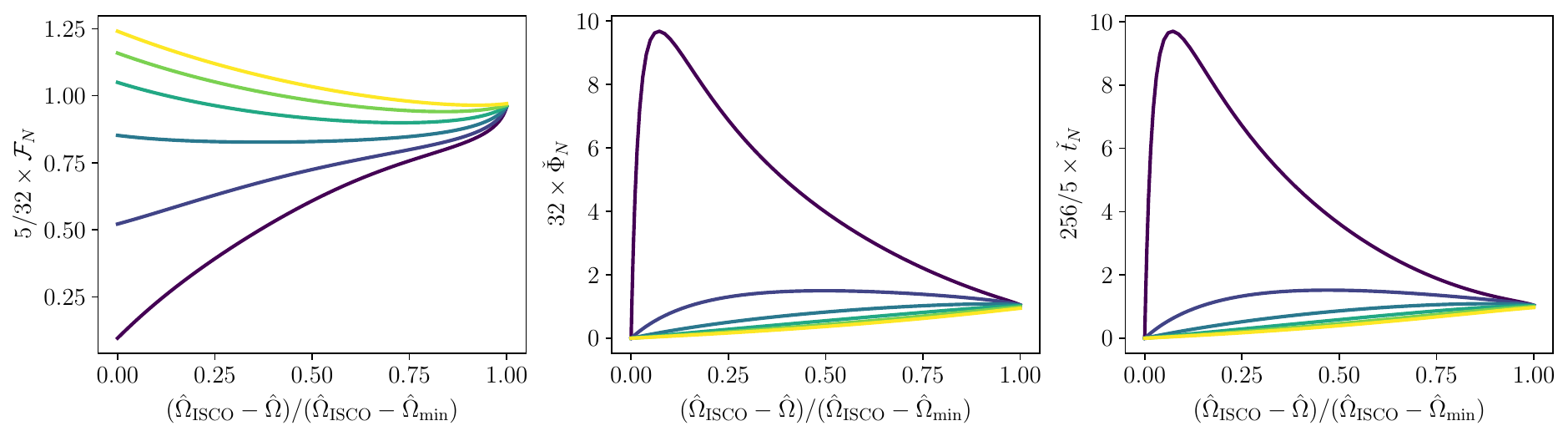}
    \caption{The normalized energy flux $\mathcal{F}_N$ (left panel), normalized orbital phase $\check{\Phi}_N$ (middle panel), and normalized orbital time to merger $\check{t}_N$ as a function of normalized distance from the ISCO frequency $\hat{\Omega}_\mathrm{ISCO}$ in frequency space. We rescale each vertical axis by the leading post-Newtonian coefficient. Thus all curves asymptote to one at low frequencies.}
    \label{fig:rescaled}
\end{figure*}

\begin{figure*}[!htp]
    \centering
    \includegraphics[width=0.98\linewidth]{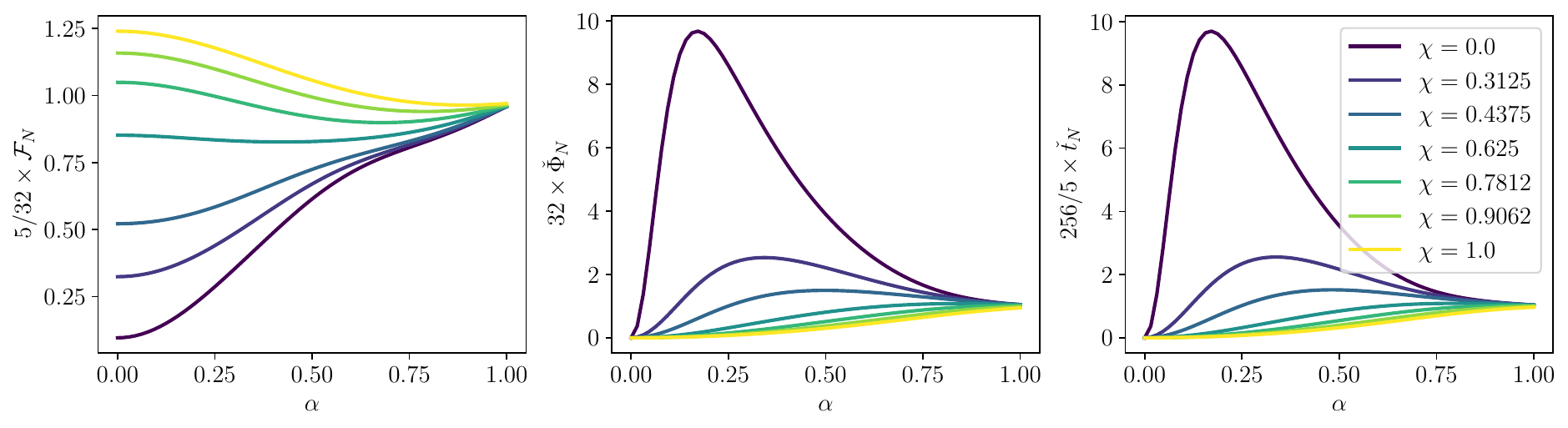}
    \caption{The normalized energy flux $\mathcal{F}_N$ (left panel), normalized orbital phase $\check{\Phi}_N$ (middle panel), and normalized orbital time to merger $\check{t}_N$ as a function of the final grid parameters $\alpha$ and $\chi$. The values of $\chi$ correspond to the same values of $\hat{a}$ used in Figure \ref{fig:original}.}
    \label{fig:final}
\end{figure*}

Consequently, we alter our parametrization to mitigate errors in our numerical spline interpolations. We first ameliorate the growth in flux, time, and phase by rescaling $\mathcal{F}_E$, $\check{t}$, and $\check{\Phi}$ by their frequency-dependence at leading post-Newtonian order,
\begin{align}
    \mathcal{F}_E^\mathrm{PN} &= \hat{\Omega}^{10/3},
    \\
    \check{\Phi}^\mathrm{PN} &= \hat{\Omega}_\mathrm{ISCO}^{-5/3} - \hat{\Omega}^{-5/3},
    \\
    \check{t}^\mathrm{PN} &= \hat{\Omega}_\mathrm{ISCO}^{-8/3} - \hat{\Omega}^{-8/3},
\end{align}
resulting in the normalized functions
\begin{align}
    \mathcal{F}_N &= \frac{\mathcal{F}_E}{\mathcal{F}_E^\mathrm{PN}},
    &
    \check{\Phi}_N &= \frac{\check{\Phi}}{ \check{\Phi}^\mathrm{PN} + \delta},
    &
    \check{t}_N &= \frac{\check{t}}{\check{t}^\mathrm{PN} + \delta},
\end{align}
which are plotted in the middle panel of Fig.~\ref{fig:rescaled}. We introduce the $\delta = 10^{-6}$ offset to avoid division by zero as $\hat{\Omega} \rightarrow \hat{\Omega}_\mathrm{ISCO}$. As a result of this shift, the normalized time and phase data still satisfy the initial conditions $\check{t}_N(\hat{\Omega}_\mathrm{ISCO}) = \check{\Phi}_N(\hat{\Omega}_\mathrm{ISCO}) = 0$. Next, motivated by post-Newtonian and near-extremal expansions of the orbital quantities, we introduce $x = \hat{\Omega}^{1/3}$ and $y = (1 - \hat{a})^{1/3}$, from which we define the final sampling parameters,
\begin{align} \label{eqn:params}
    \alpha^2 &= \frac{x_\mathrm{ISCO}-x}{x_\mathrm{ISCO}-x_\mathrm{min}},
    &
    \chi^2 &= \frac{y-y_-}{y_+-y_-},
\end{align}
where $x_\mathrm{min} = \Omega_\mathrm{min}^{1/3}$, $x_\mathrm{ISCO} = \Omega_\mathrm{ISCO}^{1/3}$, and $y_\pm = (1 \pm a_\mathrm{max})^{1/3}$. We square the left-hand side of \eqref{eqn:params} to further smooth out the behavior near the ISCO and maximal spin values. This can be seen in Fig.~\ref{fig:final}, where we plot the variation in $\check{t}$ and $\check{\Phi}$ with respect to $\alpha$ and $\chi$.

After choosing this parametrization, we solve for the fluxes $\dot{E}^\mathrm{GW}$ using the Teukolsky solver provided in the \texttt{pybhpt} Python package.\footnote{This code is available at \href{https://github.com/znasipak/pybhpt}{github.com/znasipak/pybhpt}.} Flux data is calculated with a requested precision of $10^{-10}$. We precompute $\dot{E}^\mathrm{GW}$, from which we get $\mathcal{F}_E$, on a fixed grid in the $(\alpha, \chi)$ domain using 129 equally-spaced samples in $\chi$ and 257 equally-spaced samples in $\alpha$. A downsampling of these points is mapped to the $(\hat{a}, \hat{\Omega})$ domain in Fig.~\ref{fig:domain} to illustrate the concentration of sampling points towards the ISCO and near-extremal spins. From this grid we construct the interpolated function $\mathcal{F}_N^I(\alpha, \chi)$, then solve \eqref{eqn:eom2} for $\check{t}$ and $\check{\Phi}$ on a more densely sampled $513 \times 513$ grid in $(\alpha, \chi)$. We then assemble the interpolated functions $\check{t}^I_{N}(\alpha, \chi)$ and $\check{\Phi}^I_{N}(\alpha, \chi)$. Note that we explicitly differentiate between numerical solutions and their interpolated approximants by labeling interpolated functions with the superscript $I$. Furthermore, all splines are generated by imposing the $E(3)$ boundary conditions of Behforooz and Papamichael \cite{BehfPapa79, Behf95}.

Next, we construct numerical interpolants for $\hat{\Omega}(\check{t}; \hat{a})$ and $\check{\Phi}(\check{t}; \hat{a})$. Because $\check{t}_N^I(\alpha, \chi)$ is not monotonic with respect to $\alpha$, we cannot simply parametrize $\hat{\Omega}$ and $\check{\Phi}$ in terms of $\check{t}_N$. Instead, we introduce the alternative time-parametrization
\begin{align} \label{eqn:gammaTransform}
    \gamma^6(\hat{\Omega};\hat{a}) = \ln\left[1 - {\check{t}}(\hat{\Omega};\hat{a})\right],
\end{align}
where we take the logarithm to tame the rapid growth in $\check{t}$ while preserving its monotonic relationship with $\hat{\Omega}$. The choice of $\gamma^6$ was determined through numerical experimentation, and helps to smooth out the behavior as $t\rightarrow 0$ and $\gamma \rightarrow 0$. From this we define the renormalized parameter
\begin{align} \label{eqn:betaTransform}
    \beta(\hat{\Omega}; \chi) &= \frac{\gamma(\hat{\Omega}; \hat{a}(\chi))}{\gamma_\mathrm{max}},
    &
    \gamma_\mathrm{max} &= \gamma(\hat{\Omega}_\mathrm{min};-a_\mathrm{max}),
\end{align}
and only focus on the interval $\beta \in [0, 1]$. Limiting $\beta$ does cut-off some low-frequency values in our new domain, since $\beta(\hat{\Omega}_\mathrm{min}; \chi < 1) > 1$. However, in practice this choice has a minimal impact: our frequency boundary is truncated at $\hat{\Omega} \simeq 0.00206$ instead of $\hat{\Omega} = \hat{\Omega}_\mathrm{min} = 0.002$. Note that we could extend our boundary out to $\hat{\Omega} = \hat{\Omega}_\mathrm{min}$ by redefining $\gamma_\mathrm{max}(\hat{a}) = \gamma(\hat{\Omega}_\mathrm{min}; \hat{a})$, but this functional dependence adds another layer of computational complexity when relating $\hat{\Omega}$ and $\beta$. To avoid unnecessary computational costs, we use the more simple transformation in \eqref{eqn:betaTransform}.


\begin{figure*}[!htp]
    \centering
    \includegraphics[width=0.99\linewidth]{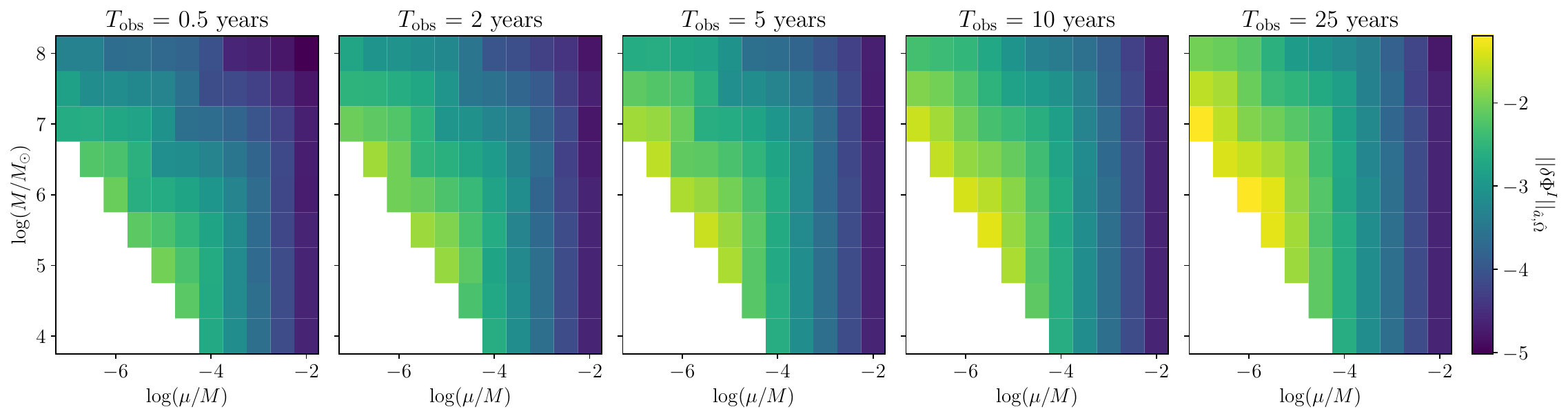}
    \caption{The logarithm of the maximized phase difference $\log||\delta \Phi^I||_{\hat{a}, \hat{\Omega}}$ as described in Sec.~\ref{sec:traj} for different observation periods $T_\mathrm{obs}$. We can see that for any observation period, all the way up to 25 years, our phase spline maintains subradian phase accuracy when compared to direct numerical integration of the orbital phase.}
    \label{fig:obsTime}
\end{figure*}

We then evaluate the time-domain equations of motion \eqref{eqn:eom1} and store solutions on an uniform $513 \times 513$ grid in $(\beta, \chi)$. Because we use $\alpha$ in place of $\hat{\Omega}$ when solving \eqref{eqn:eom1}, we must excise $\beta = 0$ ($\check{t} = 0$). The solution is trivially $\alpha = \check{\Phi} = 0$ at $\beta = 0$, but $d\alpha/dt$ diverges at this point and thus typical numerical methods for solving ordinary differential equations fail at this point. Therefore, we instead solve the equations of motion by starting one grid point away from $\beta = 0$. To incorporate the correct initial conditions, we use the Brent root-solving method to invert $\check{t}(\hat{\Omega}; a)$ and then $\check{\Phi}(\hat{\Omega}; a)$. Once we have solutions, we create the bicubic spline $\alpha^I(\beta, \chi)$, which we can transform to get $\hat{\Omega}(\check{t}; \hat{a})$. For the phase, we introduce a new renormalization that does not depend on $\hat{\Omega}$,
\begin{align} \label{eqn:phaseNorm2}
    \check{\Phi}_{N(2)} &= \ln\left(1 - \check{\Phi} \right),
\end{align}
and create the interpolant $\check{\Phi}_{N(2)}^I(\beta, \chi)$, which we can then transform to get $\check{\Phi}(\check{t}; \hat{a})$.

To evaluate the fidelity of our interpolated trajectories, we perform a series of self-consistency checks and comparisons, which are discussed in full detail in Appendix \ref{app:tests}. We find that our interpolated fluxes have fractional errors $< 10^{-8}$ over the entire domain, and they agree with independent flux calculations \cite{TaraETC14, GralHughWarb16} provided within the ``Circular Orbit Self-force Data'' repository on the Black Hole Perturbation Toolkit \cite{BHPTK18}.\footnote{Note that the flux results in the Toolkit repository are not accurate to all reported digits for $\hat{a} \geq 0.9$ and $ 2M \lesssim r_0 \lesssim 4 M$, as discussed in \ref{app:fluxtests}.} This high level of accuracy is achieved, in part, by our choice of the $E(3)$ spline boundary condition, which improves the precision of our flux interpolant by several orders of magnitude when compared to the more common `natural' or `not-a-knot' boundary conditions (see Appendix \ref{app:fluxtests}).

Furthermore, we find that the interpolated frequency-domain quantities, $\check{t}^I_N$ and $\check{\Phi}^I_N$, possess absolute errors $< 5\times 10^{-7}$, and the derivatives of the interpolated splines satisfy the frequency-domain equations of motion \eqref{eqn:eom2} to a precision $\sim 10^{-6}$. Likewise, $\alpha^I$ has estimated fractional errors $< 5\times 10^{-8}$ and $\check{\Phi}^I_{N(2)}$ has estimated absolute errors $< 5\times 10^{-8}$.

Next we quantify the impact of these interpolation errors on the gravitational wave phases. Recall from \eqref{eqn:adiabaticWaveform} that the phase of each gravitational wave mode $h_{\ell m}$ is proportional to the orbital phase $\Phi$. Therefore, the accuracy of the gravitational wave phases is set by the accuracy of $\Phi$. To quantify this accuracy, we first calculate the orbital phase accumulated over an inspiral with initial orbital frequency $\hat{\Omega}_0$ using our interpolated data,
\begin{align}
    \Delta \Phi^I &= \frac{1}{\epsilon}\left[\check{\Phi}^I\left(\check{t}_0 + \frac{\epsilon T}{M}; \hat{a} \right) - \check{\Phi}^I\left(\check{t}_0; \hat{a}\right) \right],
\end{align}
where $\check{t}_0 = \check{t}(\hat{\Omega}_0; \hat{a})$ and $T$ is the duration of the observed inspiral. Note that $\check{\Phi}^I$ is reconstructed from $\check{\Phi}^I_{N(2)}$ via \eqref{eqn:phaseNorm2}. We then compare $\Delta \Phi^I$ to the accumulated orbital phase $\Phi^\mathrm{ODE}$ obtained by directly integrating \eqref{eqn:eom1} with the initial conditions $\hat{\Omega}(\check{t}=0) = \hat{\Omega}_0$ and $\check{\Phi}(\check{t}=0) = 0$. The interpolation error is then estimated by the difference $\delta \Phi^I = |\Delta\Phi^I - \Phi^\mathrm{ODE}|$. Next, we find the maximum value of $\delta \Phi^I$ for a range of $\hat{a}$ values and initial orbital frequencies $\hat{\Omega}_0$. We denote this maximized difference as $||\delta \Phi^I||_{\hat{a},\hat{\Omega}}$. In Figure \ref{fig:obsTime} we plot $||\delta \Phi^I||_{\hat{a},\hat{\Omega}}$ as a function of mass-ratio $10^{-7} \leq \epsilon \leq 10^{-2}$ and massive black hole mass $10^{4} \leq M \leq 10^{8}$ for binaries observed for $T = [0.5, 2, 5, 10, 25]$ years. We exclude binaries with secondary masses $\mu < M_\odot$. We measure maximum phase errors of $0.0097$, $0.0186$, $0.0307$, $0.0430$, and $0.0642$ for the observation periods $0.5$, $2$, $5$, $10$, and $25$ years, respectively. These phase errors are reduced by about a factor of two if we only focus on inspirals with initial frequencies $\hat{\Omega} \geq 0.013$ (corresponding to $r_0 \lesssim 18 M)$ and binaries with smaller bodies of mass $\mu \geq 2 M_\odot$. Because the $(\ell, m) = (2,2)$ mode dominates each gravitational wave signal, $2 \Delta\Phi^I$ provides a reasonable estimate of the gravitational wave phase error. Based on this approximation, our interpolation errors most likely meet the subradian gravitational wave phase accuracy requirements for realistic space-based mHz gravitational wave observations, which we further verify in Sec.~\ref{sec:errors}.

\subsection{Mode amplitudes}
\label{sec:amp}

Similar to the trajectories, we presample the complex waveform amplitudes $B_{\ell m}$ on a fixed $65 \times 65$ grid in $(\alpha, \chi)$. To optimize the accuracy of our amplitude interpolation, we decompose $B_{\ell m}$ into its phase $\psi_{\ell m}$ and the log of its real amplitude $\ln A_{\ell m}$. We then interpolate these quantities independently. We generate mode data for $\ell \leq 15$ and $0 < m \leq \ell$. Even with the reduced sampling, the interpolation errors introduce estimated fractional errors $< 3\times 10^{-5}$ in $A_{\ell m}$ and absolute errors $<2 \times 10^{-5}$ for $\psi_{\ell m}$, as discussed in Appendix \ref{app:tests}.

\begin{figure}[!tp]
    \centering
    \includegraphics[width=0.95\linewidth]{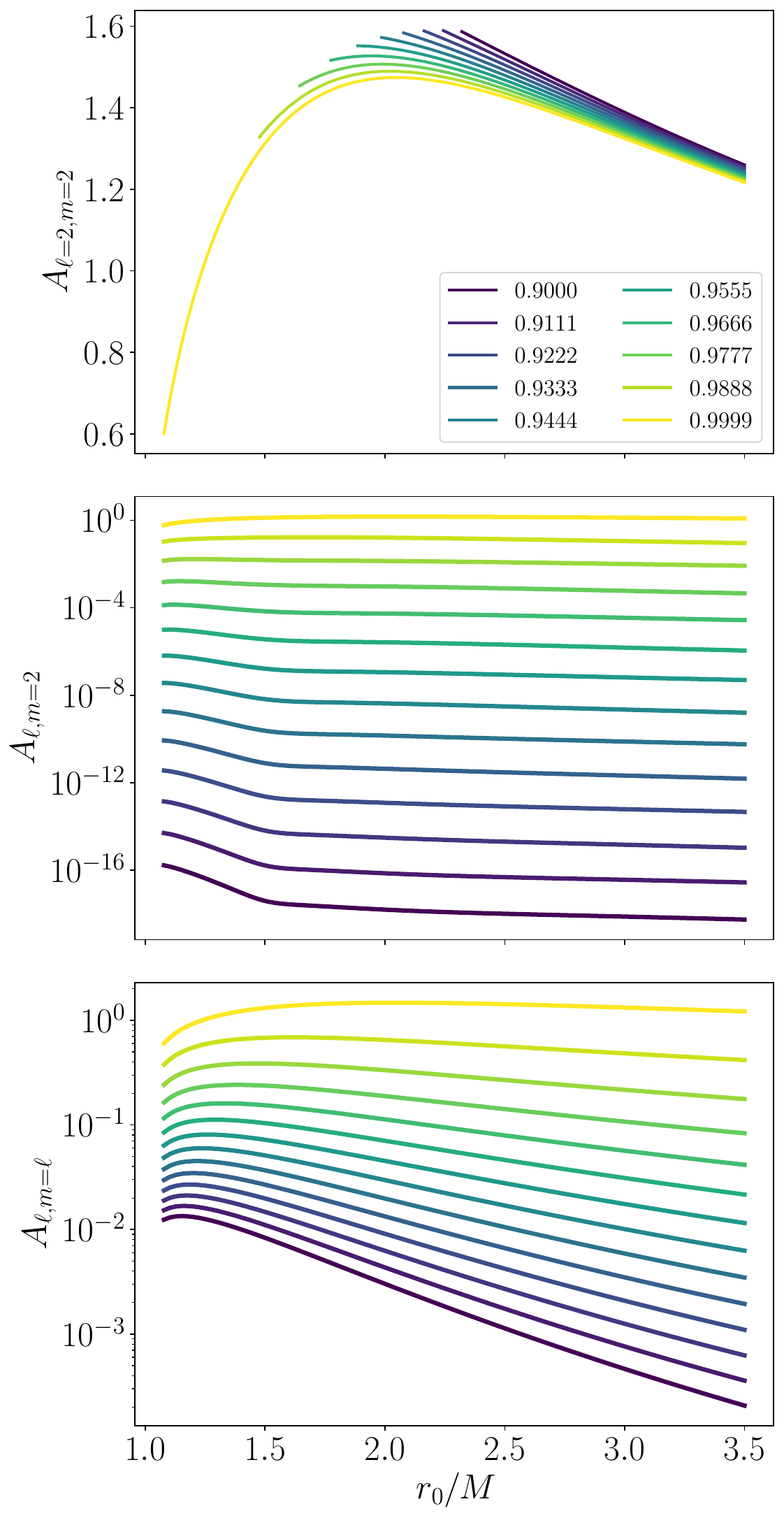}
    \caption{Waveform amplitudes $A_{\ell, m}$ as a function of the radius of the orbit $r_0/M$. The top panel plots $A_{\ell=2, m=2}$ for different Kerr spin parameters $\hat{a}\geq 0.9$. The middle and bottom panels plot $A_{\ell, m=2}$ and $A_{\ell, m=\ell}$, respectively, for $\hat{a} = 0.9999$ and $2 \leq \ell \leq 15$. In the two lower panels, the mode magnitudes are ordered inversely with $\ell$. In other words, the $\ell = 2$ mode is the dominant mode, the $\ell = 3$ is the next most dominant mode, then $\ell =4$, $\ell = 5$, and so on, with $\ell = 15$ being the mode with the smallest magnitude (and darkest solid line). }
    \label{fig:amplitudeVariation}
\end{figure}

The amplitudes display a number of interesting properties, particularly as we move into the near-horizon extremal Kerr (NHEK) regime, i.e., $1-\hat{a}^2 \ll 1$ and $(r-r_+)/r_+ \ll 1$. For $\hat{a} \lesssim 0.995$ the amplitudes peak at the ISCO and decay with the orbital frequency. For $\hat{a} \gtrsim 0.995$ the amplitudes can instead peak well before reaching the ISCO and then decay as the orbital frequency increases. This behavior has been reported in previous investigations of extremal Kerr black holes \cite{GralPorfWarb15, GralHughWarb16}. For the dominant $\ell = m = 2$ mode, the waveform amplitudes reach their maximum magnitudes near the stationary surface of the outer ergosphere $r_E^+ = 2M$, as demonstrated in the top panel of Figure \ref{fig:amplitudeVariation}. The modes do not have the same turnover behavior for $\ell \neq m$, but instead plateau around $r_E^+$ before rising again as they approach the ISCO, which is highlighted in the middle panel of Figure \ref{fig:amplitudeVariation} for all $m =2$ modes. This behavior is due to our reprojection of the $(j, m)$-modes of a spin-weighted spheroidal harmonic basis to the $(\ell, m)$-modes of a spherical basis, given in \eqref{eqn:XfromR}. As demonstrated in \cite{GralPorfWarb15}, larger spheroidal $j$ modes (for a fixed $m$) will decay more rapidly as one approaches the ISCO in the NHEK regime. Therefore, in the spherical $(\ell, m)$-basis, the near-ISCO behavior is dominated by the contribution from the modes around $j=m$, even when $\ell \gg m$. As a result, for $\ell > m$ the spherical-spheroidal mixing with many $j\sim m$ modes prevents the turnover and decay of $A_{\ell m}$ as $\Omega \rightarrow \Omega_\mathrm{ISCO}$.

Finally, we consider the relative power between the $\ell = m = 2$ and $\ell = m = 15$ modes for a given orbital frequency. In the bottom panel of Figure \ref{fig:amplitudeVariation}, we plot the $A_{\ell, m = \ell}$ amplitudes as functions of $r_0$ for different values of $\hat{a}$. We see that $|A_{15,15}/A_{2,2}|^2 \lesssim 5\times 10^{-4}$ for all orbital frequencies and spins, placing a conservative upper bound for the relative power contributed by the $\ell = m = 15$ mode at a fixed frequency. For most year-long EMRI signals, the relative \emph{total} power (i.e., the power integrated across the entire frequency evolution of an inspiral) contributed by the $\ell = m = 15$ mode is $\lesssim 10^{-6}$, as we demonstrate in the following section. Therefore, we find it unnecessary to include higher harmonic contributions beyond $\ell \leq 15$.

\subsection{Waveform evaluation and mode selection}
\label{sec:wave}

Given a fixed time step $dt$, signal duration $T$, and binary source parameters (e.g., $M$, $\mu$, $r_0$, $d_L$), time-domain waveforms are constructed using \eqref{eqn:adiabaticWaveform}, \eqref{eqn:SSBtoSource}, \eqref{eqn:polarization}, with all functions evaluated from the trajectory and mode amplitude splines outlined above. We simplify the sum over $(\ell, m)$-modes using $B_{\ell -m} = (-1)^\ell B^*_{\ell m}$ and the identities of spin-weighted spherical harmonics,
\begin{subequations}
    \begin{align}
    Y_{s\ell -m}(\theta) &= (-1)^{m} Y_{-s \ell m}(\theta),
    \\
    &= (-1)^{\ell} Y_{s \ell m}(\pi-\theta),
    \end{align}
\end{subequations}
leading to the reduced sums
\begin{subequations} \label{eqn:timeDomainReduced}
    \begin{align}
    h_+(t) = +\sum_{\ell=2}^\infty \sum_{ m = 1}^{\ell} \mathcal{A}_{\ell m} Y^+_{\ell m} \cos(\psi_{\ell m} + m[\phi - \Phi]),
    \\
    h_\times(t) = -\sum_{\ell=2}^\infty \sum_{ m = 1}^{\ell}  \mathcal{A}_{\ell m} Y^\times_{\ell m} \sin(\psi_{\ell m} + m[\phi - \Phi]),
\end{align}
\end{subequations}
where $\mathcal{A}_{\ell m} = \mu A_{\ell m}/r$ and
\begin{subequations}
    \begin{align}
    Y^{+/\times}_{\ell m}(\theta) &= Y_{-2\ell m}(\theta) \pm Y_{-2\ell m}(\pi - \theta),
    \\
    &= Y_{-2\ell m}(\theta) \pm (-1)^{\ell + m} Y_{+2\ell m}(\theta).
    \end{align}
\end{subequations}

The truncation of the $(\ell, m)$-mode sum is determined by the power in each harmonic mode,
\begin{subequations} \label{eqn:powerLM}
    \begin{align}
        P_{\ell m} &= \int_0^T \left[ \left|\mathcal{A}_{\ell m}(t) Y_{\ell m}^+\right|^2 + \left|\mathcal{A}_{\ell m}(t) Y_{\ell m}^\times\right|^2  \right]dt ,
        \\
        &\approx \left[\left|Y_{\ell m}^+\right|^2 + \left|Y_{\ell m}^\times\right|^2 \right] \sum_{n=1}^{N} |\mathcal{A}_{\ell m}(\hat{\Omega}_n)|^2 \Delta t_n,
    \end{align}
\end{subequations}
where $\hat{\Omega}_n = \hat{\Omega}(t=0) + n \Delta\hat{\Omega}$, with orbital frequency spacing $\Delta \hat{\Omega} = [\hat{\Omega}(t=T) - \hat{\Omega}(t=0)]/N$, time spacing $\Delta t_n = t(\hat{\Omega}_n) - t(\hat{\Omega}_{n-1})$, and signal duration $T$. In this work, we find that $N = 500$ sufficiently approximates the mode power. Additionally, we find that an equal spacing in frequency space is more numerically efficient than an equal sampling in time. We include all $(\ell, m)$-modes that satisfy the selection criteria
\begin{align} \label{eqn:modeSelectionCriteria}
    {P_{\ell m}} > \epsilon_\mathrm{mode} \times {\sum_{\ell' = 2}^{\ell} \sum_{m'=1}^{\ell'}P_{\ell' m'}},
\end{align}
for a user-specified threshold $\epsilon_\mathrm{mode}$. Rather than computing the power for all of the modes and then removing the modes that do not satisfy \eqref{eqn:modeSelectionCriteria}, we instead perform a serial search. First we find all modes that meet \eqref{eqn:modeSelectionCriteria} for $m \leq 2$. We then increase $m$ to $m = 3$ and increment over $\ell$, beginning with $\ell = m$ until \eqref{eqn:modeSelectionCriteria} is no longer satisfied. We repeat this process of increasing $m$ and then $\ell$, until either $P_{mm}$ no longer satisfies \eqref{eqn:modeSelectionCriteria} or $m > \ell_\mathrm{max} = 15$.

In Table \ref{tab:modeCount}, we report the number of selected modes $N_\mathrm{mode}$ for a variety of inspirals and threshold values. As expected, varying $\epsilon_\mathrm{mode}$ has the most significant impact on the number of modes included. Additionally, increasing the large mass-ratio $M/\mu$ or the primary mass $M$ enhances $N_\mathrm{mode}$, most likely because these changes also increase the amount of time that binary spends orbiting near the ISCO, where subdominant modes are the most pronounced. Likewise, subdominant modes are more significant at smaller separations, which is why increasing $\hat{a}$ also leads to more modes being selected. On the other hand, increasing the duration of the waveform decreases $N_\mathrm{mode}$, since more power comes from earlier in the inspiral, where the subdominant modes are much weaker.

\begin{figure*}[!htp]
    \centering
    \includegraphics[width=0.98\linewidth]{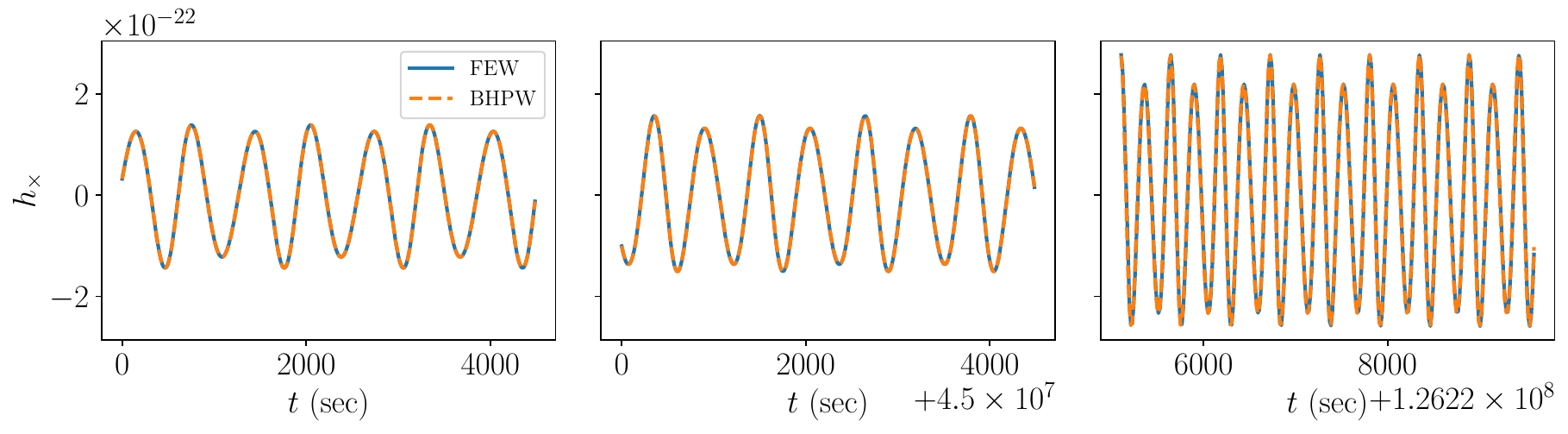}
    \includegraphics[width=0.98\linewidth]{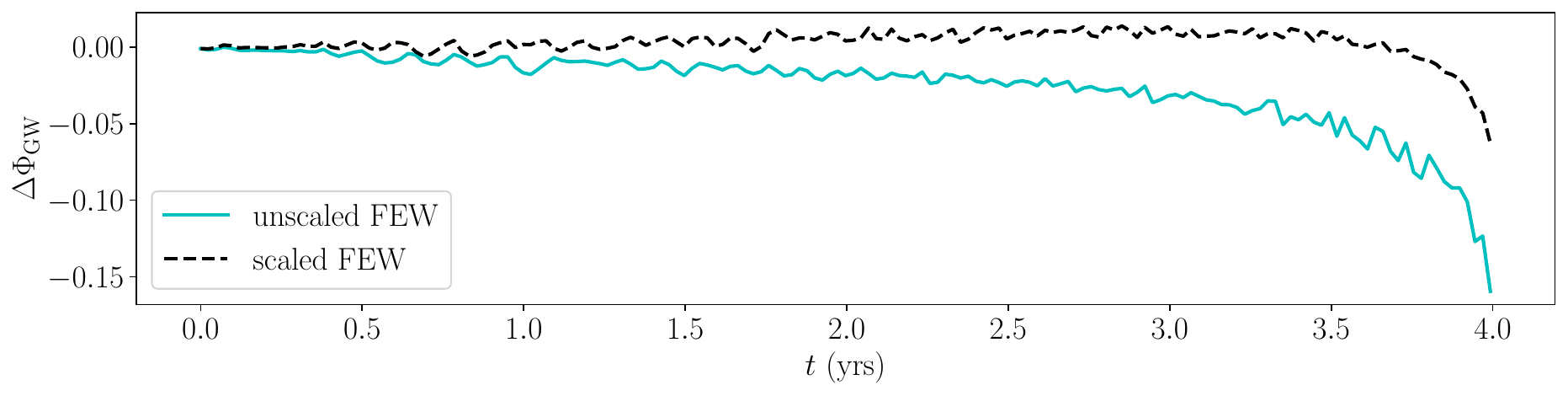}
    \caption{Comparisons between waveforms produced by the FEW and \texttt{bhpwave} models for an EMRI with intrinsic parameters $(M, \mu, a, p_0, e_0, x) = (10^6 M_\odot, 10 M_\odot, 0, 12.05, 0, 1)$. The top panels show the evolution of the strain $h_+$ with time $t$. In the top plots, the solid (blue) lines correspond to the waveform computed by FEW, while the dashed (orange) lines are produced by \texttt{bhpwave}. The three different panels focus on three small time windows within the full four year signal. The bottom panel plots the difference in the gravitational wave phases between the waveform models. The solid (cyan) line shows the phase difference when the waveforms are evaluated at the same set of parameters. The dashed (black) line is the phase difference after rescaling the FEW parameters to account for different definitions of $GM_\odot$.}
    \label{fig:compareFEW}
\end{figure*}

In the final column of Table \ref{tab:modeCount} we also report the relative power of the $\ell = m = 15$ mode, $P_{15,15}/P_\mathrm{tot}$, where $P_\mathrm{tot}$ is the summed power from all selected $(\ell, m)$-modes. Consequently, this mode is only included if $\epsilon_\mathrm{mode} \leq P_{15,15}/P_\mathrm{tot}$. Previous investigations of adiabatic model suggest that $ 10^{-2} \lesssim \epsilon_\mathrm{mode} \lesssim 10^{-5}$ is a sufficient threshold range to prevent systematic biases in EMRI data analysis for LISA \cite{KatzETC20}. Thus, we see that there is no need to go beyond the $\ell = 15$ modes for the range of spin values considered by our model.

\renewcommand{\arraystretch}{1.5}
\begin{table}
    \caption{The number of selected $(\ell, m)$-modes $N_\mathrm{mode}$ for a given inspiral $(\hat{a}, M, \mu, T)$ and mode selection threshold $\epsilon_\mathrm{mode}$. The initial conditions are chosen so that each system reaches the ISCO after $T$ years. In the final column we also include the relative power in the $\ell = m = 15$ mode, $P_{15,15}/P_\mathrm{tot}$, where $P_\mathrm{tot}$ is the summed power from all selected $(\ell, m)$-modes.}
    \label{tab:modeCount}
    \centering
    \begin{tabular*}{\linewidth}{c @{\extracolsep{\fill}} cccccc}
        \hline
        \hline
         $\hat{a}$ & $M \,(M_\odot)$ & $\mu \,(M_\odot)$ & $T$ (yrs) & $\epsilon_\mathrm{mode}$ & $N_\mathrm{mode}$ & $\frac{P_{15,15}}{P_\mathrm{tot}}$
         \\
         \hline
         $0.5000$ & $10^6$ & $10$ & $0.1$ & $10^{-5}$ & 14 & $2\times10^{-9}$
         \\
         $0.9000$ & $10^6$ & $10$ & $0.1$ & $10^{-5}$ & 18 & $1\times10^{-7}$
         \\
         $0.9950$ & $10^6$ & $10$ & $0.1$ & $10^{-5}$ & 24 & $3\times10^{-6}$
         \\
         \hline
         $0.9999$ & $10^5$ & $10$ & $0.1$ & $10^{-5}$ & 17 & $8\times10^{-7}$
         \\
         $0.9999$ & $10^6$ & $10$ & $0.1$ & $10^{-5}$ & 28 & $1\times10^{-5}$
         \\
         $0.9999$ & $10^7$ & $10$ & $0.1$ & $10^{-5}$ & 39 & $2\times10^{-4}$
         \\
         \hline
         $0.9500$ & $10^5$ & $10$ & $1.0$ & $10^{-5}$ & 11 & $5\times10^{-9}$
         \\
         $0.9500$ & $10^6$ & $10$ & $1.0$ & $10^{-5}$ & 15 & $6\times10^{-8}$
         \\
         $0.9500$ & $10^7$ & $10$ & $1.0$ & $10^{-5}$ & 23 & $1\times10^{-6}$
         \\
         \hline
         $0.9999$ & $10^6$ & $10$ & $0.5$ & $10^{-5}$ & 23 & $4\times10^{-6}$
         \\
         $0.9999$ & $10^6$ & $10$ & $2.0$ & $10^{-5}$ & 20 & $2\times10^{-6}$
         \\
         $0.9999$ & $10^6$ & $10$ & $4.0$ & $10^{-5}$ & 18 & $1\times10^{-6}$
         \\
         \hline
         $0.9999$ & $10^6$ & $10$ & $0.1$ & $10^{-3}$ & 10 & $1\times10^{-5}$
         \\
         $0.9999$ & $10^6$ & $10$ & $0.1$ & $10^{-4}$ & 16 & $1\times10^{-5}$
         \\
         $0.9999$ & $10^6$ & $10$ & $0.1$ & $10^{-6}$ & 40 & $1\times10^{-5}$
         \\
         \hline
         \hline
    \end{tabular*}
\end{table}

Following this mode selection, we evaluate all of the selected modes in \eqref{eqn:timeDomainReduced} at the time steps $t_i = i\times dt$ for $i = 0, 1, 2, \dots, T/dt$. To speed-up the calculation, we parallelize evaluations over the time steps $t_i$ using OpenMP \cite{OpenMP}.

Frequency-domain waveforms are generated in a similar manner for a fixed frequency step $df$ and maximum frequency $f_\mathrm{max}$. Alternatively, one can specify a time step $dt$ and signal duration $T$, for which we set $df = 1/T$ and $2f_\mathrm{max} = 1/dt$. By eliminating sums over negative $m$-modes, we have
\begin{align}
    \tilde{h}_+(f) &= \frac{1}{2}\sum_{\ell = 2}^\infty \sum_{m = 1}^{\ell} \sqrt{\frac{2\pi}{m\dot{\Omega}}}\mathcal{A}_{\ell m} Y_{\ell m}^+ e^{+i\Psi_{\ell m}},
    \\
    \tilde{h}_\times(f) &= \frac{i}{2}\sum_{\ell = 2}^\infty \sum_{m = 1}^{\ell} \sqrt{\frac{2\pi}{m\dot{\Omega}}}\mathcal{A}_{\ell m} Y_{\ell m}^\times e^{+i\Psi_{\ell m}},
\end{align}
where the phases are given by $\Psi_{\ell m}(f) = \psi_{\ell m}(\hat{\Omega}_f/m) + 2\pi f t(\hat{\Omega}_f/m) + m[\phi - \Phi(\hat{\Omega}_f/m)] - \pi/4$ and all mode-functions are evaluated at $\hat{\Omega}_f = 2\pi f$. Mode selection is identical to the time-domain waveforms. Similarly we parallelize evaluations over the frequency steps $f_j = j\times df$ for $j = 0, 1, 2, \dots, f_\mathrm{max}/df$.

As a final note, waveforms produced by retrograde orbits can be parametrized one of two ways: (1) by keeping the massive black hole spin positive and setting the orbital angular momentum to be negative, ($a \geq 0$ and $x_0 \leq 0$); or (2) vice versa ($a \leq 0$ and $x_0 \geq 0$). These two parametrizations are identical up to the transformation $(\theta, \phi) \rightarrow (\pi - \theta, -\phi)$, as shown in Appendix \ref{app:retrograde}. In this work, we keep $x_0 = 1$ fixed at its positive value and vary the sign of $a$, allowing for a much smoother transition from prograde to retrograde orbits in the equatorial limit. This choice also has the advantage of keeping all other orbital constants (e.g, $\Omega_p$, $\mathcal{L}_z$) strictly positive. However, users can still specify $x_0 = -1$, and internally we construct the waveform using
\begin{align}
    h(t; a, -x_0, \theta, \phi) &= h(t; -a, x_0, \pi - \theta, -\phi).
\end{align}
For waveforms in the SSB frame, $\phi$ is fixed, while the orientation of the spin vector is set by $q_K$ and $\phi_K$. Therefore we introduce the azimuthal shift through $\Phi_{\phi 0}$ and transform $\theta$ via a parity inversion of $(q_K, \phi_K)$, leading to
\begin{align}
    &h_\mathrm{SSB}(t; a, -x_0, q_K, \phi_K, \Phi_{\phi 0}) =
    \\
    \notag
    &\qquad h_\mathrm{SSB}(t; -a, x_0, \pi - q_K, \pi + \phi_K, \pi + \Phi_{\phi 0}).
\end{align}
The transformations are identical for the frequency-domain waveforms.

\subsection{Model comparison}
\label{sec:few}

To demonstrate the accuracy of our model, we compare our waveforms with those produced by FEW for $\hat{a} = e_0 = 0$. As a visual demonstration, in Figure \ref{fig:compareFEW} we plot $h_+$ for an EMRI with source properties $(M, \mu, p_0) = (10^6 M_\odot, 10 M_\odot, 12.05)$ observed in the SSB frame for four years. The strain calculated by FEW is given by the dashed lines, while the \texttt{bhpwave} result is given by the solid lines. The waveforms are aligned to agree at time $t=0$. We zoom in on the waveform at three time windows placed near the beginning, middle, and end of the waveform. Across all three periods there is minimal disagreement between the two models. In the bottom panel of Figure \ref{fig:compareFEW}, we also plot the phase difference between the two models as a function of time. The two waveforms maintain subradian agreement over all five years of the signal.

The agreement between the two models is further improved when we account for the fact that \texttt{bhpwave} and FEW use different values of $G M_\odot$, which impacts unit conversions as one goes from the geometric units ($G=c=1$) in the inspiral calculation to SI units for the observed waveform. In \texttt{bhpwave}, we use the gravitational parameter $G M_\odot = 1.32712440041279419 \times 10^{20}\, \mathrm{m}^3\, \mathrm{s}^{-2}$ from Jet Propulsion Laboratory's (JPL) planetary and lunar ephemerides \cite{ParkETC21}. The fractional difference between this value and the FEW value is $1.557\times 10^{-8}$. If we rescale time and mass quantities by this difference, we improve the phase agreement between FEW and \texttt{bhpwave} by a factor of $\sim 5$, as demonstrated by the dashed line in the right panel of Figure \ref{fig:compareFEW}.

For a more quantitative comparison, in Table \ref{tab:compareFEW} we report the mismatch between the \texttt{bhpwave} waveforms $h_\mathrm{B}$ and the FEW waveforms $h_\mathrm{F}$ for a series of different EMRI systems. The mismatch is defined by
\begin{align}
    \mathcal{M}(h_\mathrm{B}, h_\mathrm{F}) = 1 - \sum_{+,\times} \frac{\left(h_\mathrm{B} | h_\mathrm{F} \right)}{\sqrt{\left(h_\mathrm{B} | h_\mathrm{B} \right)\left(h_\mathrm{F} | h_\mathrm{F} \right)}},
\end{align}
where
\begin{align}
    \left(a | b \right) = 4 \mathrm{Re} \int_0^\infty \frac{\tilde{a}(f)\tilde{b}^*(f)}{S_n(f)}df,
\end{align}
is the noise-weighted inner product between two real signals, $S_n(f)$ is the one-sided noise spectral density of LISA, and the sum is performed over the plus and cross polarizations of the gravitational wave strain. For this work, we use the analytic approximation of $S_n(f)$ given in \cite{RobsCornLiu19}, and provided through the \texttt{lisatools} Python package.\footnote{\href{https://github.com/mikekatz04/LISAanalysistools}{https://github.com/mikekatz04/LISAanalysistools}} We estimate the Fourier transforms of our time-domain signals by first applying a tapered cosine window, also known as a Tukey window, and then taking the discrete Fourier transform (DFT) of the windowed function. We choose the tapered cosine window shape parameter $\alpha = 0.001$, which leads to minimal loss in the signal-to-noise ratio (SNR) $\rho = \sqrt{(h|h)}$, while also providing an improved estimate of the Fourier transform. We find mismatches $\lesssim 10^{-3}$ across different sets of intrinsic source parameters.

\begin{figure}[!htp]
    \centering
    \includegraphics[width=0.98\linewidth]{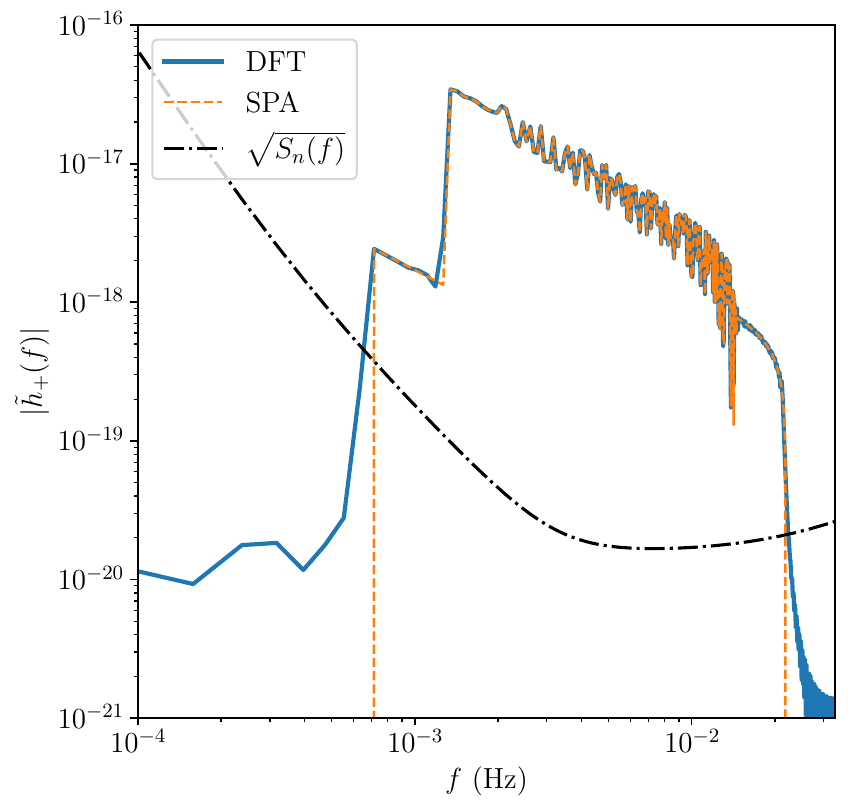}
    \caption{Comparison between the discrete Fourier transform of the time-domain gravitational wave strain $h_+$ produced by \texttt{bhpwave} (solid blue line) and the frequency-domain waveform for $h_+$ produced by \texttt{bhpwave} using the stationary phase approximation (dashed orange line). For reference, the sensitivity curve $S_n$ is also plotted as the dot-dashed (black) curve. Note that we downsample the number of the frequency samples included in this plot by a factor of 10000 in order to more clearly see how well the two waveforms agree, even in the highly oscillatory region between $10^{-3} - 10^{-2}$ Hz.}
    \label{fig:dft}
\end{figure}

\renewcommand{\arraystretch}{1.5}
\begin{table}
    \caption{Mismatches between \texttt{bhpwave} waveforms and FEW waveforms for different sets of intrinsic parameters $\vec{\theta}_\mathrm{intrinsic} = \{M, \mu, p_0, \Phi_{\phi0}= 0.2\}$ and signal duration $T$. We scale the distance so that each signal has an SNR $\rho = 20$. The third to last column reports the mismatch between \texttt{bhpwave} and FEW when both models are evaluated at the same set of parameters $\vec{\theta}_0$. In the second to last column, we give an estimate of the ``best-fit'' mismatch between the two waveform models. This is computed by evaluating \texttt{bhpwave} at $h_\mathrm{B}(\vec{\theta}_0)$ and FEW at $h_\mathrm{F}(\vec{\theta}_0- \Delta \vec{\theta}_0)$, where $ \Delta \vec{\theta}_0$ is determined using the Cutler-Vallisneri bias estimate. The maximum ratio between the systematic biases and the statistical errors estimated from the \texttt{bhpwave} model,  $\mathcal{R} = ||\Delta \vec{\theta}_0/\Delta \vec{\theta}_\mathrm{stat}||_\infty$, is reported in the last column. }
    \label{tab:compareFEW}
    \centering
    \begin{tabular*}{\linewidth}{c @{\extracolsep{\fill}} cccccc}
        \hline
        \hline
         $M/M_\odot$ & $\mu/M_\odot$ & $p_0$ & $T/$yrs &  $\mathcal{M}$ & $\mathcal{M}_\mathrm{bf}$ & $\mathcal{R}$
         \\
         \hline
         $10^5$ & $\phantom{0}1$ & $15.9$ & $1.55$ & $4\times 10^{-3}$ & $2\times 10^{-5}$ & $0.12$
         \\
         $10^5$ & $10$ & $15.9$ & $0.16$ & $6\times 10^{-5}$ & $1\times 10^{-5}$ & $0.01$
         \\
         $10^6$ & $\phantom{0}1$ & $15.9$ & $8.00$ & $7\times 10^{-4}$ & $3\times 10^{-5}$ & $0.11$
         \\
         $10^6$ & $10$ & $15.9$ & $8.00$ & $5\times 10^{-4}$ & $8\times 10^{-5}$ & $0.12$
         \\
         $10^6$ & $10$ & $12.0$ & $4.00$ & $8\times 10^{-3}$ & $6\times 10^{-5}$ & $0.07$
         \\
         $10^6$ & $50$ & $15.9$ & $3.00$ & $6\times 10^{-4}$ & $6\times 10^{-5}$ & $0.16$
         \\
         $10^7$ & $50$ & $10.0$ & $8.00$ & $9\times 10^{-5}$ & $5\times 10^{-5}$ & $0.07$
         \\
         $10^7$ & $50$ & $\phantom{0}7.5$ & $2.60$ & $3\times 10^{-4}$ & $1\times 10^{-5}$ & $0.04$
         \\
         \hline
         \hline
    \end{tabular*}
\end{table}

Furthermore, we perform a self-consistency check between our time-domain and frequency-domain waveforms. In Figure \ref{fig:dft} we plot the frequency-domain waveform $\tilde{h}_+(f)$ for the system $(M, \mu, a/M, p_0) = (10^6M_\odot, 30 M_\odot, 0.9, 13.55)$. The system is observed for $T=4$ years (but merges after $\sim 3.95$ years) at a time step of $dt=15$ seconds, which corresponds to $f_\mathrm{max} = 1/30$ Hz and $df \simeq 1.6\times 10^{-8}$ Hz. Overlaying the DFT of the time-domain waveform $\mathrm{DFT}[h_+](f)$ for the same system, we find a good overlap between the two signals, which have a mismatch of $\mathcal{M} = 6\times 10^{-5}$. However, as seen in other investigations of the frequency-domain EMRI waveforms \cite{SperETC23}, the mismatch is highly dependent on the sample size due to the spectral leakage inherent in the DFT. For example the mismatch increases to $\mathcal{M} = 7\times 10^{-2}$ for $dt=10$ seconds or to $\mathcal{M} = 3\times 10^{-3}$ for $dt=2$ seconds. Note that the agreement can also be improved by windowing the time-series data \cite{SperETC23}.









\section{Assessing modeling errors}
\label{sec:errors}

Finally, we use \texttt{bhpwave} to provide a few examples of how we can assess the impact of modeling errors and systematics on EMRI parameter estimation. In Bayesian inference, the probability that a given set of model parameters $\vec{\theta}$ describes an observed signal $d(t)$ is given by the posterior distribution
\begin{align}
    p(\theta | d) \propto p(d|\theta) p(\theta),
\end{align}
where the prior distribution $p(\theta)$ is the probability of observing the parameters $\vec{\theta}$ (in the absence of any signal or evidence) and the likelihood $p(\theta |d)$ is the probability of the evidence $d(t)$ given fixed parameters $\vec{\theta}$. In gravitational wave data analysis, the (log) likelihood reduces to \cite{Whit53}
\begin{align} \label{eqn:ll}
    \log p(d|\theta) \propto -\frac{1}{2}(d - h_m(\theta)| d- h_m(\theta)),
\end{align}
for a waveform model $h_m$. When assuming uniform priors, the likelihood provides an unnormalized estimate for the posterior distribution. Thus \eqref{eqn:ll} peaks at the model parameters $\vec{\theta}_\mathrm{peak}$ that ``best fit" the data, and the width of this peak corresponds the statistical certainty $\Delta \vec{\theta}_\mathrm{stat}$ with which we can measure $\vec{\theta}_\mathrm{peak}$.

Naturally, every model possesses some degree of numerical or systematic error, which can bias $\vec{\theta}_\mathrm{peak}$ away from the true parameters $\vec{\theta}_\mathrm{true}$ that describe $d(t)$. Consequently, a waveform model is considered sufficiently accurate for parameter estimation if the systematic biases $\Delta \vec{\theta}_\mathrm{bias} = \vec{\theta}_\mathrm{peak} - \vec{\theta}_\mathrm{true}$ are smaller than the inherent statistical uncertainty $\Delta \vec{\theta}_\mathrm{stat}$ in the measurement.

We can estimate this intrinsic statistical uncertainty via
\begin{align} \label{eqn:stat}
    \Delta {\theta}^{i}_\mathrm{stat} \approx \sqrt{\left[\Gamma^{-1}_m(\theta_\mathrm{peak})\right]^{ii}},
\end{align}
where $\Gamma^{-1}_m$ is the inverse of the Fisher information matrix,
\begin{align} \label{eqn:fisher}
    \Gamma^{ij}_m(\theta) = \left( {\partial_i h_m(\theta_\mathrm{peak})} \vert {\partial_j h_m(\theta_\mathrm{peak})} \right),
\end{align}
and $\partial_i = \partial/\partial \theta^i$.
Likewise the impact of the systematic errors can be estimated from the Cutler-Vallisneri bias \cite{CutlVall07}
\begin{align} \label{eqn:bias}
    \Delta {\theta}^{i}_\mathrm{bias} \approx \left[\Gamma^{-1}_m(\theta_\mathrm{peak})\right]^{ij} \left(\partial_j h_m(\theta_\mathrm{peak}) \vert \Delta h_m(\theta_\mathrm{peak}) \right),
\end{align}
where $\Delta h_m(\theta_\mathrm{peak}) = h_t(\theta_\mathrm{peak}) - h_m(\theta_\mathrm{peak})$ is the difference between the ``true'' strain $h_t$ and the biased strain $h_m$ produced by our model.

\begin{figure}[!htp]
    \centering
    \includegraphics[width=0.95\linewidth]{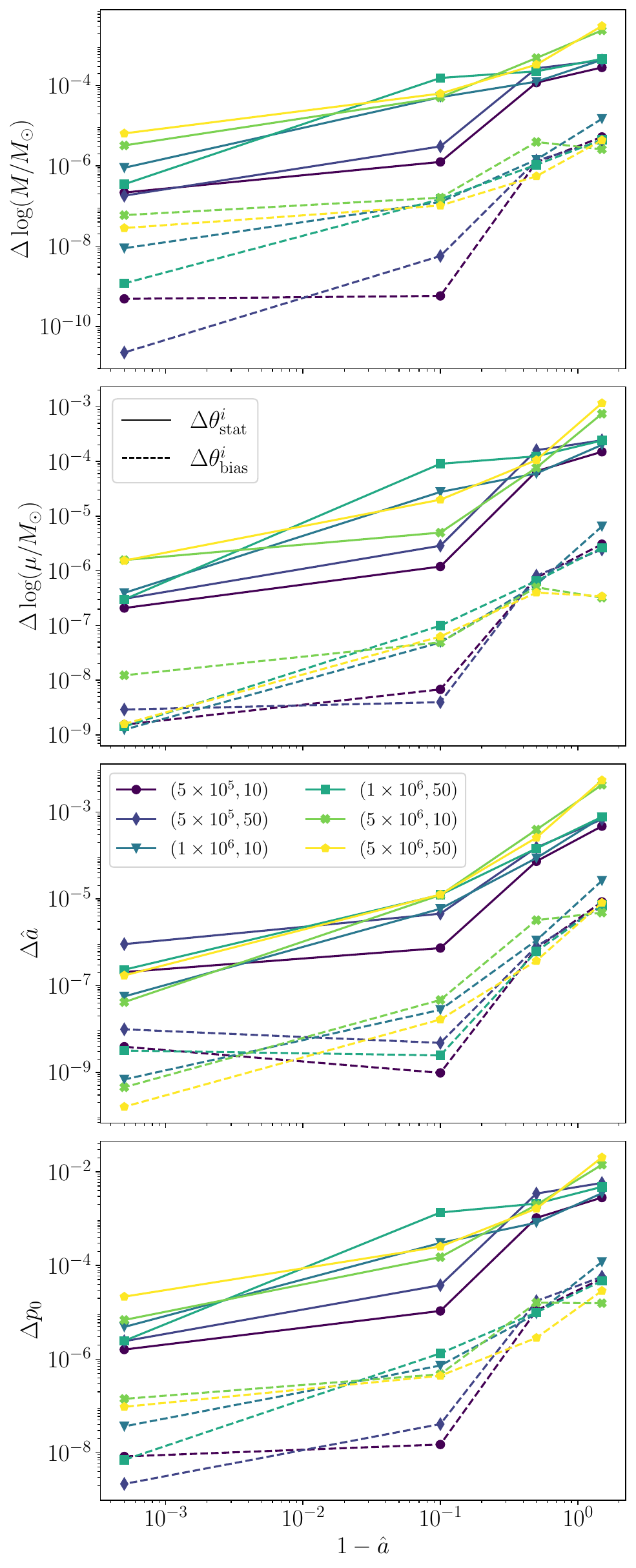}
    \caption{Comparisons of the statistical uncertainties $\Delta\theta^i_\mathrm{stat}$ (solid lines) and systematic biases $\Delta\theta^i_\mathrm{bias}$ (dashed lines) in the intrinsic parameters $(M,\mu,\hat{a},p_0)$ as a function of spin $\hat{a}$ for a range of binaries with different masses. Different mass combinations $(M,\mu)$ corresponds to different colors and markers, as given by the legend in the third panel. The systematic biases are introduced by downsampling the trajectory data to introduce larger interpolation errors into our waveform model.}
    \label{fig:fisherCompare}
\end{figure}

\begin{figure*}[!htp]
    \centering
    \includegraphics[width=0.98\linewidth]{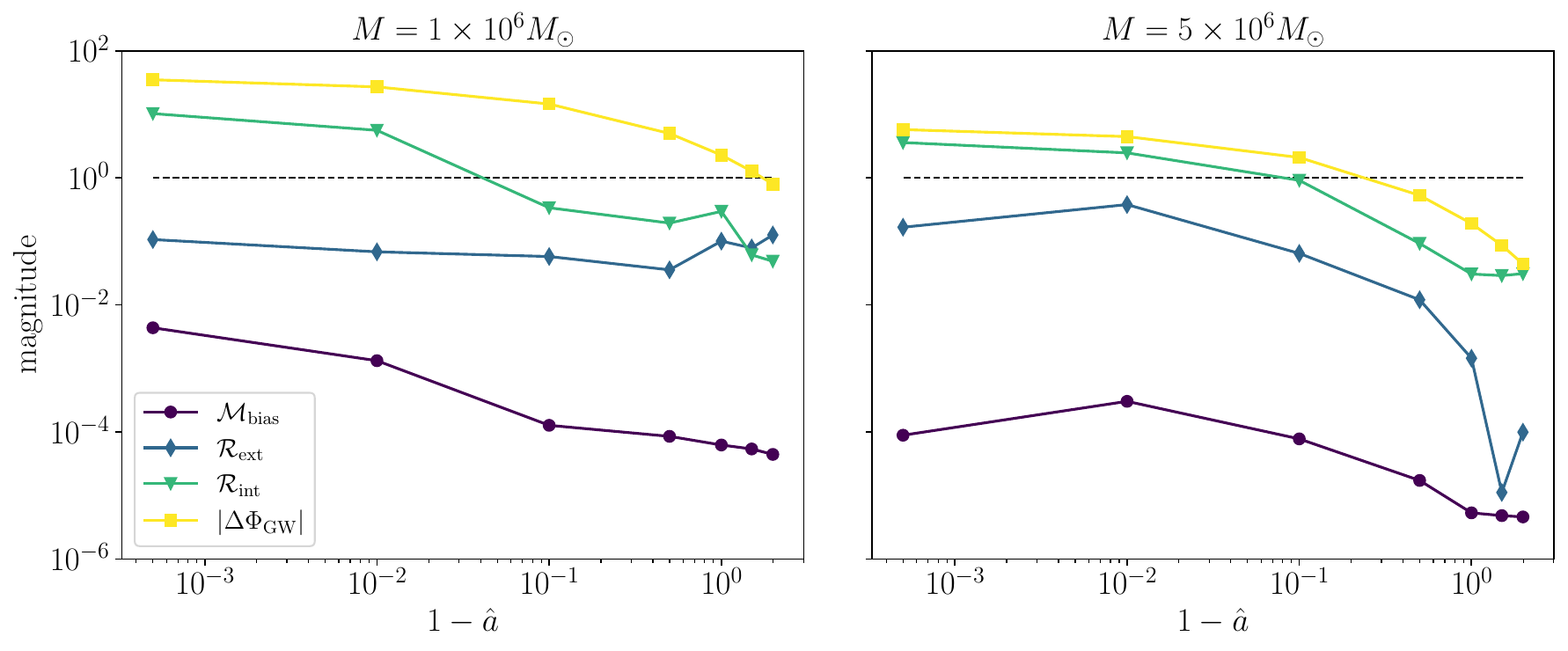}
    \caption{The impact of flux errors on parameter estimation as a function of spin $\hat{a}$, as discussed in Sec.~\ref{sec:errors}. The circular markers (purple line) plot the mismatch between $h_E(\theta)$ and $h_B(\theta-\theta_\mathrm{bias})$, where $h_B$ is our accurate model, $h_E$ is our biased model, $\theta$ are the parameters of the GW source predicted by $h_B$ and $\theta_\mathrm{bias}$ are the biases in those parameters from the true values $\theta-\theta_\mathrm{bias}$. The diamond markers (blue line) plot the maximum ratio between these biases and the intrinsic uncertainty in the extrinsic parameters, while the triangle markers (green line) plot the same ratio for the intrinsic parameters. The square markers (yellow line) report the overall dephasing between the two models. The dashed line marks a magnitude of one.}
    \label{fig:errorBiases}
\end{figure*}

First we verify that the mismatches between the \texttt{bhpwave} and FEW models correspond to small biases, such that $\Delta {\theta}^{i}_\mathrm{bias} < \Delta {\theta}^{i}_\mathrm{stat}$. To do this, we take FEW to be the ``true" waveform $h_t = h_\mathrm{F}$ and \texttt{bhpwave} to be the model waveform $h_m = h_\mathrm{B}$. Next, we solve \eqref{eqn:stat} and \eqref{eqn:bias} at the ``peak" parameters listed in Table \ref{tab:compareFEW}. Appendix \ref{app:fisher} details our numerical procedures for constructing and verifying the Fisher matrix and its inverse. We then calculate the maximum ratio between the systematic biases and statistical uncertainties across all of the parameters,
\begin{align}
    \mathcal{R} = \bigg|\bigg|\frac{\Delta \theta^i_\mathrm{bias}}{\Delta {\theta}^i_\mathrm{stat}}\bigg|\bigg|_\infty,
\end{align}
which are reported in the last column of Table \ref{tab:compareFEW}. We find that $\mathcal{R} \lesssim 0.1$ across all of the considered systems, indicating that the systematic errors between the two waveform models will not bias parameter estimates in this region of parameter space (i.e., non-eccentric, non-spinning EMRIs). Furthermore, to verify the accuracy of our bias estimates, we calculate the ``true" parameters $\theta^i_\mathrm{true} = \theta^i_\mathrm{peak} - \Delta \theta^i_\mathrm{bias}$ and then compute the mismatch between $h_\mathrm{F}(\theta_\mathrm{true})$ and $h_\mathrm{B}(\theta_\mathrm{peak})$. As shown in the second to last column of Table \ref{tab:compareFEW}, correcting for the biases between the two models reduces their mismatch across all of the tabulated source parameters.

Next, we consider how phase errors introduce systematic biases to our waveform model. To quantify this, we take the time-domain phase and frequency splines, $\check{\Phi}(\check{t}; a)$ and $\check{\Omega}(\check{t}; a)$, and downsample their interpolated data sets by a factor of two in each dimension. This introduces a known source of interpolation noise into our waveform model. The (noisy) waveforms generated by this downsampled data are denoted by $h_\mathrm{D}$. We then take the original \texttt{bhpwave} waveform generator to be the truth (e.g., $h_t = h_\mathrm{B}$), $h_\mathrm{D}$ to be our model (e.g., $h_m = h_\mathrm{D}$), and evaluate \eqref{eqn:stat} and \eqref{eqn:bias} for peak parameters $\theta^i_\mathrm{peak} = (M, \mu, a, p_0, e_0, x_0, d_L, q_S, \phi_S, q_K, \phi_K, \Phi_{\phi0}, \Phi_{\theta0}, \Phi_{r0})$. For this analysis we hold $(e_0 = 0, x_0 = 1, q_S = 0.3, \phi_S = 1.3, q_K = 1.8, \phi_K = 1.2, \Phi_{\phi 0} = 0.2, \Phi_{\theta0} = 0, \Phi_{r0} = 0)$ constant and vary over $M/M_\odot = \{5\times 10^{5}, 10^{6}, 5\times 10^{6}\}$, $\mu/M_\odot = \{10, 50\}$, and $a = \{-0.5, 0.5, 0.9, 0.995\}$. Each signal is observed for $T = 2$ years at a time spacing of $dt = 10 $ seconds until it reaches the ISCO. This sets the value of $p_0$. Additionally, we scale the distance $d_L$ so that each signal has a SNR $\rho = 30$. We find that $\mathcal{R} < 0.1$ for all of the sources.

In Figure \ref{fig:fisherCompare} we plot the resulting statistical uncertainties and systematic biases of $(\Delta\log(M/M_\odot), \Delta\log(\mu/M_\odot), \Delta \hat{a}, \Delta p_0)$ as a function of $1-\hat{a}$ for the various sources. Solid lines refer to the statistical uncertainties, while dashed lines refer to the systematic biases. Different colors refer to different combinations of $M$ and $\mu$. We see that the systematic biases always fall under the intrinsic uncertainty estimates. Furthermore, the uncertainties decrease as spin increases, with the biases possessing largely the same behavior. Thus, at least in the region of parameter space considered for this toy analysis, we see that the interpolation errors in the phases and frequencies of our adiabatic trajectories do not significantly bias our waveforms.

As a final test, we inject artificial noise into our original flux data and then recompute all of the trajectory splines to create a new biased waveform model, which we denote by $h_E$. The flux is modified via the replacement
\begin{align}
    \mathcal{F}_E \rightarrow\mathcal{F}^\delta_E = \frac{1 + (1 + \delta)\frac{1247}{336}\hat{\Omega}^{2/3}}{1 + \frac{1247}{336}\hat{\Omega}^{2/3}} \mathcal{F}_E,
\end{align}
where we set $\delta = 10^{-5}$. With this rescaling, the post-Newtonian expansion of $\mathcal{F}^\delta_E$ is
\begin{align}
    \mathcal{F}^\delta_E \sim \frac{32}{5}\hat{\Omega}^{10/3}\left[1 - \frac{1247}{336}(1 - \delta) \hat{\Omega}^{2/3} + O(\hat{\Omega}) \right].
\end{align}
By setting $\delta = 0$ we recover the first two post-Newtonian orders of the energy flux (e.g., see (178) in \cite{SasaTago03}), while setting $\delta = 10^{-5}$ adds a slight perturbation at the first (subleading) post-Newtonian order.
From a post-adiabatic standpoint, this error is on the same order as 1PA corrections for systems with mass ratios $\mu/M \sim \delta$.

Once again, we take the original \texttt{bhpwave} model to represent the true waveform and the corrupted model to represent our model waveform. We then compute $\Delta \theta^i_\mathrm{stat}$ and $\Delta \theta^i_\mathrm{bias}$ for a variety of systems. As before we fix $(e_0 = 0, x_0 = 1, q_S = 0.3, \phi_S = 1.3, q_K = 1.8, \phi_K = 1.2, \Phi_{\phi 0} = 0.2, \Phi_{\theta0} = 0, \Phi_{r0} = 0)$ and use the same signal duration, $p_0$ values, and time step size. For this analysis we only consider the masses $(M, \mu) = (10^6 M_\odot, 10 M_\odot)$ and $(M, \mu) = (5\times 10^6 M_\odot, 10 M_\odot)$, and we vary over the spins $a = \{-0.99, -0.5, 0., 0.5, 0.9, 0.99, 0.9995\}$.

In Figure \ref{fig:errorBiases} we plot the dephasing $\Delta \Phi_\mathrm{GW} = 2(\Phi_B - \Phi_E)$ of the two gravitational wave models, the maximum ratio between $\Delta \theta^i_\mathrm{stat}$ and $\Delta \theta^i_\mathrm{bias}$ for the intrinsic parameters $\mathcal{R}_\mathrm{intrinsic}$, the maximum ratio between $\Delta \theta^i_\mathrm{stat}$ and $\Delta \theta^i_\mathrm{bias}$ for the extrinisic parameters $\mathcal{R}_\mathrm{extrinsic}$, along with the mismatch $\mathcal{M}_\mathrm{bf}$ between $h_B(\theta-\Delta \theta_\mathrm{bias})$ and $h_E(\theta)$. Note that we estimate $\Delta \Phi_\mathrm{GW}$ from the dephasing between just the $(2,2)$-modes of the two waveforms, which is two times the difference between the phase trajectories of the \texttt{bhpwave} and corrupted models, $\Phi_B$ and $ \Phi_D$, respectively. The left panel includes results for the binary with masses $(M, \mu) = (10^6 M_\odot, 10 M_\odot)$, while the right panel is for a binary with $(M, \mu) = (5\times 10^6 M_\odot, 10 M_\odot)$. The dashed line denotes a magnitude of 1. Thus, dephasings below the dashed line indicate subradian agreement between the two models, and ratios below the line indicate that the systematic biases are below the intrinsic uncertainty of the measured parameters.

For both systems we see that the dephasing and biases in the intrinsic parameters tend to increase with spin. This is expected, since the error scales with the frequency, and higher spin systems will reach larger orbital frequencies and consequently possess larger errors. However, large dephasings do not necessarily guarantee that the systematic errors will significantly bias the measured parameters. For instance, in the $(M, \hat{a})=(10^6 M_\odot, 0.9)$ system, $\Delta \Phi_\mathrm{GW} \sim 10$ radians, yet the systematic biases remain smaller than the statistical parameter uncertainties. This result mirrors recent studies of EMRI waveforms of systems with non-rotating massive black hole, but with post-adiabatic effects and spinning secondaries included \cite{BurkETC23}.

Furthermore, for both systems, the biases in the extrinsic parameters are consistently below the parameter uncertainties. At least in this simplified quasi-circular case, this suggests that errors in the trajectory predominantly affect the intrinsic properties of the source. Interestingly, after accounting for the biases, we still generate small mismatches between the models across all of the tested systems. This indicates that our erroneous models, despite their large biases, can still capture most of the power in the EMRI gravitational wave signal (in the absence of noise). Thus, adiabatic waveforms may be sufficiently accurate for measuring the extrinsic parameters of quasi-circular EMRIs in the LISA data stream and systems with very low spins or retrograde orbits. It remains to be seen if this would also hold true in the case of eccentric orbits or if we were to incorporate a realistic LISA response in our analysis.

\section{Conclusion}
\label{sec:conclusion}

We presented the theoretical and numerical methods behind \texttt{bhpwave}: a new Python-based adiabatic gravitational waveform generator for binary sources composed of a compact object undergoing a quasi-circular inspiral into a rotating massive black hole. To build this waveform model, we precomputed mass-independent trajectories for systems with initial separations $r_0/M \lesssim 60$ and Kerr spins $|a| \leq 0.9999$. Furthermore, we precomputed all potential waveform harmonic amplitudes for $\ell \leq 15$. By implementing the $E(3)$ boundary conditions in our numerical spline algorithm, we improved the precision of our interpolated flux data and trajectories by as much as three orders of magnitude over traditional spline methods. We computed waveforms in both the time and frequency domains and observed good agreement between these models. Furthermore, we compared our model against the FEW waveform generator for $\hat{a} = e_0 = 0$ and achieved mismatches $\mathcal{M} \sim 10^{-5}$. Using Fisher matrix calculations, we also assessed the magnitudes of any systematic biases introduced by numerical error in our model. We found that biases due to interpolation error are well below the thresholds required for LISA data analysis. Additionally, we demonstrated that waveform dephasing does not provide a complete picture of modeling error. In particular, for systems with retrograde orbits and slowly-rotating massive black holes, we found that waveform models could have phase errors of up to 10 radians, and yet the biases introduced by these errors would not be measurable by LISA. Thus, the claim that waveform models require subradian phase accuracy for LISA is a useful guide for model fidelity, but more sophisticated analyses are required to understand the true impact of modeling errors on LISA science (see also \cite{BurkETC23}).

A notable limitation of \texttt{bhpwave} is that it currently neglects the effects of eccentricity and precession. Nonetheless, while most observed EMRIs are expected to be highly eccentric \cite{GairETC04},  there are possible formation channels driven by accretion flow that circularize EMRI dynamics and align the orbital angular momentum with the massive black hole's spin \cite{PanLyuYang21}. Thus, \texttt{bhpwave} is applicable to these so-called ``wet-formation" EMRIs.

In the future, we plan on integrating our Kerr data in the FEW model in order to leverage its ability to run on GPUs. Additionally, moving forward we will use \texttt{bhpwave} to perform a more thorough investigation of how different interpolation schemes, levels of flux accuracy, trajectory parametrizations, and mode selection criteria can impact LISA data analysis. In particular, we plan to incorporate a second-generation time-delay interferometry (TDI) response and verify Fisher matrix calculations by performing full MCMC samplings (similar to the one provided in Appendix \ref{app:fisher}). With a more detailed and rigorous investigation of these computational systematics, we can put more stringent bounds on the accuracy requirements for EMRI waveform modeling, which will be particularly important as we design more complicated (eccentric, precessing) EMRI waveform models.

\begin{acknowledgements}
This research was supported by an appointment to the NASA Postdoctoral Program at the NASA Goddard Space Flight Center, administered by Oak Ridge Associated Universities under contract with NASA. The author thanks Michael Katz and Lorenzo Speri for extensive discussions and advice on EMRI data analysis, as well as John G.~Baker and Benjamin Leather for their insight and helpful feedback. This work makes use of the Black Hole Perturbation Toolkit. It also uses the \texttt{Eryn}\footnote{\href{https://github.com/mikekatz04/Eryn}{https://github.com/mikekatz04/Eryn}} and \texttt{lisatools}\footnote{\href{https://github.com/mikekatz04/LISAanalysistools}{https://github.com/mikekatz04/LISAanalysistools}} repositories authored by Michael Katz.
\end{acknowledgements}

\appendix

\section{Approximations of the frequency domain waveforms}
\label{app:fourierPhase}

In the small mass-ratio limit, $M\dot{\Phi} = M\Omega \sim O(1)$ while $M \dot{\psi}_{\ell m} \sim O(\epsilon)$. Therefore, neglecting the time-evolution of ${\psi}_{\ell m}$, we invert $f\approx m\Omega(t)/(2\pi)$ to approximate $t_p(f)$ for each $(\ell, m)$ mode. At first glance, one might expect that ignoring this $O(\epsilon)$ term would introduce an error of $O(1)$ in the values of $t_p$ and $\Phi(t_p)$, thus diminishing the phase accuracy of our frequency-domain waveform. However, these errors perfectly cancel, leading to an $O(\epsilon)$ error. This can be seen by parametrizing the time and phase in terms of $\Omega$. Then $\Omega_f = \Omega(f) = \Omega_0 + \delta\Omega$ where $\Omega_0 = 2\pi f/m$ and $\delta\Omega \sim O(\epsilon)$. The induced error in the phasing $\Psi(f) = 2\pi f t(\Omega_f) + \psi_{\ell m}(\Omega_f) - m\Phi(\Omega_f)$ for a fixed value of $f$ is then given by
\begin{align}
    \Psi(f) &=
    m\Omega_0 t(\Omega_0) + \psi_{\ell m}(\Omega_0) - m\Phi(\Omega_0)
    \\ \notag
    & \quad \qquad + \big[m\Omega_0 \partial_\Omega t(\Omega_0) + \partial_\Omega\psi_{\ell m}(\Omega_0)
    \\ \notag
    & \quad \qquad \qquad - m\partial_\Omega\Phi(\Omega_0)\big]\delta \Omega + O(\delta\Omega^2),
    \\
    &=
    m\Omega_0 t(\Omega_0) + \psi_{\ell m}(\Omega_0) - m\Phi(\Omega_0)
    \\ \notag
    & \quad \qquad \qquad + \partial_\Omega\psi_{\ell m}(\Omega_0) \delta \Omega + O(\delta\Omega^2),
\end{align}
where we have made use of the fact that
$\Omega \partial_\Omega t = \partial_\Omega \Phi$. Finally, we take into account that $\partial_\Omega \psi_{\ell m} = \dot{\psi}_{\ell m}\partial_\Omega t$. Since $M^2 \dot{\Omega} \sim O(\epsilon)$, then $\partial_\Omega \psi_{\ell m} \, \delta \Omega \sim O(\epsilon)$, which sets the overall error in the phase at $O(\epsilon)$ due to neglecting $\dot{\psi}_{\ell m}$.

Furthermore, when calculating $\tilde{B}_{\ell m}(f)$, we neglect any contribution from $\ddot{\psi}_{\ell m}$ in \eqref{eqn:fourierAmplitude}. We expect this approximation to introduce an $O(\epsilon)$ error relative to the leading-order behavior of $\tilde{B}_{\ell m}(f) \sim 1/\sqrt{\epsilon}$, and therefore is safe to neglect in the small mass-ratio limit.

\section{Validating fluxes, trajectories, and numerical splines}
\label{app:tests}

We describe several validation tests for assessing the accuracy of our inspiral trajectory data.

\subsection{Numerical precision of fluxes}
\label{app:fluxtests}

\begin{figure}[!tp]
    \centering
    \includegraphics[width=0.95\linewidth]{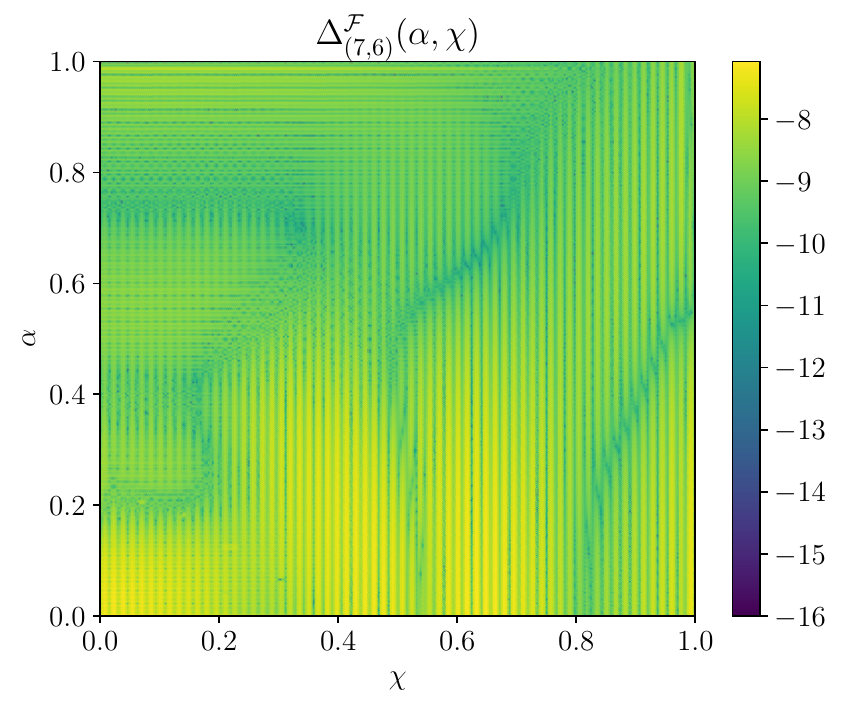}
    \caption{The fractional interpolation error $\Delta^\mathcal{F}_{(7,6)}$ of the flux spline $\mathcal{F}_{N(7,6)}^I$ as a function of the grid parameters $\alpha$ and $\chi$.}
    \label{fig:fluxSplineError}
\end{figure}

\begin{figure}[!t]
    \centering
    \includegraphics[width=0.95\linewidth]{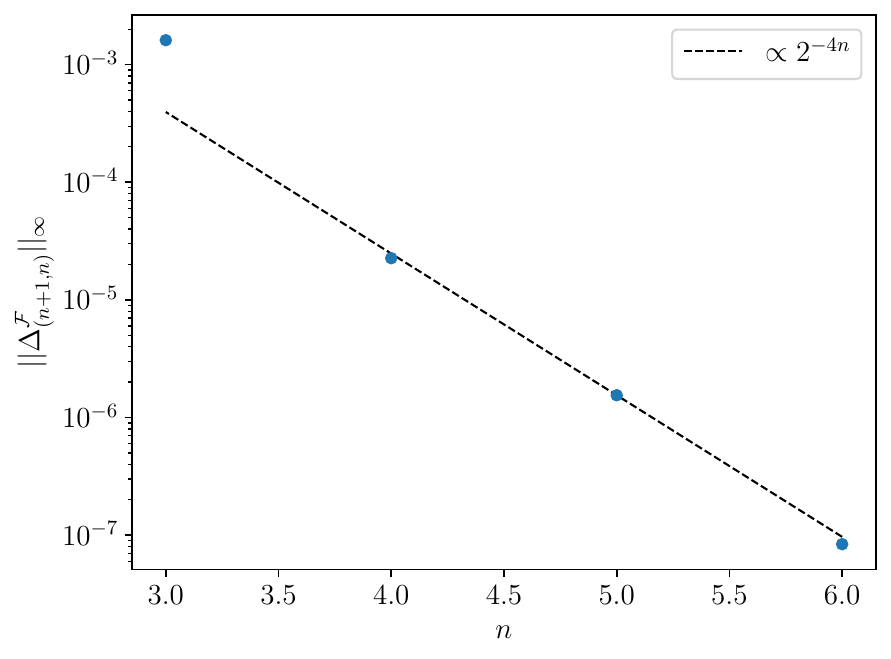}
    \caption{Convergence of the maximum fractional interpolation error $||\Delta^\mathcal{F}_{(n+1,n)}||_\infty$ as we increase $n$. Increasing $n$ by one amounts to  doubling the sampling density in each dimension of our grid. Thus a power law decay of $2^{-4n}$ (dashed blue line) indicates fourth-order convergence of our bicubic spline.}
    \label{fig:fluxSplineConvergence}
\end{figure}

\begin{figure*}[bhtp]
    \centering
    \includegraphics[width=0.95\linewidth]{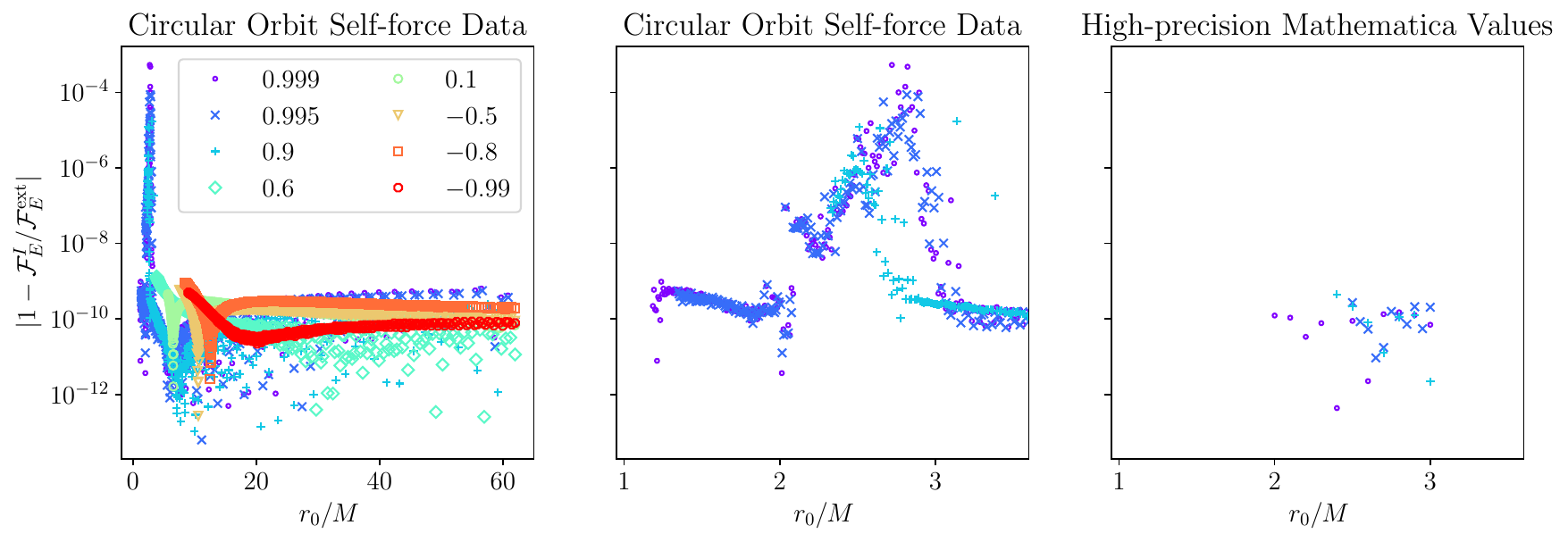}
    \caption{Fractional error between our interpolated flux data $\mathcal{F}_E^I$ and the flux data reported within the Circular Orbit Self-force Data repository in the BHPToolkit. The left panel displays errors across the entire domain covered by our spline. The middle plot focuses on a region where we find strong disagreement between our model and the repository. The right plot compares our data to flux calculations from a high-precision ($>100$ digits) Mathematica code in this problem region, demonstrating that our results are consistent with high-precision calculations.}
    \label{fig:fluxComparison}
\end{figure*}

\begin{figure}[bhtp]
    \centering
    \includegraphics[width=0.95\linewidth]{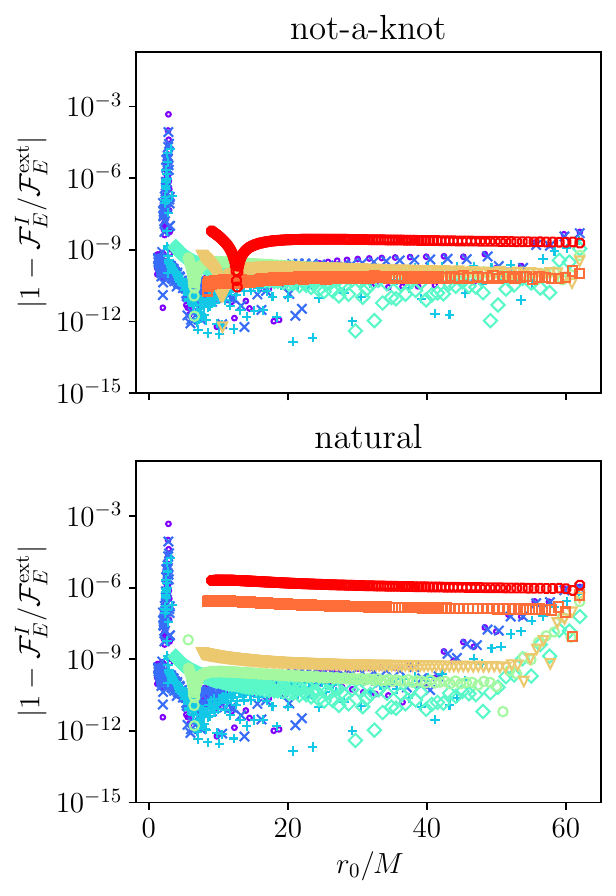}
    \caption{The same comparison as the left plot of Figure \ref{fig:fluxComparison}, only this time we construct our flux interpolant by imposing not-a-knot boundary conditions (top panel) or natural boundary conditions (bottom panel), leading to worse agreement with the pretabulated flux data.}
    \label{fig:fluxComparison2}
\end{figure}

To confirm the numerical accuracy of our interpolated data, we perform a series of comparisons and self-consistency checks. First we assess the accuracy of our flux interpolant by downsampling the data. Let $f^I_{(n,m)}(x,y)$ denote a bicubic spline that approximates a function $f(x,y)$ by interpolating sampled values of $f$ on a rectilinear $(2^n+1) \times (2^m+1)$ grid in $(x,y)$. Thus, in this notation our fully-sampled interpolant $\mathcal{F}_N^I(\alpha, \chi)$ can also be expressed as $\mathcal{F}_{N(8,7)}^I(\alpha, \chi)$, because it is constructed from a $257\times 128$ grid of flux values in $(\alpha, \chi)$. We then estimate the fractional interpolation error for the flux spline $\mathcal{F}_{N(N',M')}^I$ via
\begin{align}
    \Delta^\mathcal{F}_{(N',M')}(\alpha, \chi) = \left|1 - \frac{\mathcal{F}_{N(N',M')}^I(\alpha, \chi)}{\mathcal{F}_{N(N'+1,M'+1)}^I(\alpha, \chi)} \right|.
\end{align}
By downsampling the data, we calculate $\Delta^\mathcal{F}_{(7,6)}$, $\Delta^\mathcal{F}_{(6,5)}$, and $\Delta^\mathcal{F}_{(5,4)}$. We plot $\Delta^\mathcal{F}_{(7,6)}$ in Figure \ref{fig:fluxSplineError}, demonstrating that $\Delta^\mathcal{F}_{(7,6)}<10^{-6}$ across the entire domain. To estimate the interpolation error of our fully-sampled spline, $\Delta^\mathcal{F}_{(8,7)}$, we examine the convergence rate of $|| \Delta^\mathcal{F}_{(n+1,n)} ||_\infty$ in Figure \ref{fig:fluxSplineConvergence}. We find that $|| \Delta^\mathcal{F}_{(n+1,n)} ||_\infty \propto 2^{-4n}$, which is consistent with the standard fourth-order scaling for cubic spline errors (error $ \sim \Delta x^4$ for grid spacing $\Delta x$). Provided this scaling holds true as we increase the sampling rate, we estimate that $\Delta^\mathcal{F}_{(8,7)} \lesssim 5\times 10^{-8}$.

We also compare the flux interpolant to Kerr circular flux values produced by independent codes \cite{TaraETC14, GralHughWarb16}. Tables of these values are provided within the ``Circular Orbit Self-force Data'' repository hosted by the Black Hole Perturbation Toolkit \cite{BHPTK18}. Figure \ref{fig:fluxComparison} plots the fractional errors between the Toolkit data set and our interpolated fluxes for the spin values $\hat{a} = [-0.99, -0.8, -0.5, 0.1, 0.6, 0.9, 0.995, 0.999]$ as a function of the orbital separation $r_0$. For $r_0 \gtrsim 3.5 M$ the fractional errors in the fluxes are consistent with the interpolation error estimated in Figure \ref{fig:fluxSplineError}, indicating that our flux results are reliable to a precision $\sim 10^{-9}$ in this domain. Crucially, we find that our use of the $E(3)$ boundary condition in our spline interpolation is essential for achieving these small fractional errors. For example, in Figure \ref{fig:fluxComparison2} we plot the fractional errors resulting from the use of the more common `natural' or `not-a-knot' spline boundary conditions when interpolating our flux results. Both boundary conditions degrade the accuracy of our interpolated flux data, with the natural spline providing the largest fractional errors. For both the natural and not-a-knot splines, the fractional errors rise as we approach larger negative values of the spin ($a \rightarrow -1$) and larger orbital separations ($\Omega \rightarrow \Omega_\mathrm{min}$), where our flux data is sparsely-sampled. Therefore, a careful choice of boundary conditions can significantly improve the accuracy of our splines, thereby reducing the number of points at which we need to perform expensive calculations of the gravitational wave fluxes.

Finally, we return the near-ISCO region of Figure \ref{fig:fluxComparison}. The fractional errors between our data and the fluxes published in the Toolkit peak at $\sim 10^{-3}$ near $r_0 \sim 3 M$, as seen in the middle plot of Figure \ref{fig:fluxComparison}. To identify the source of this disagreement, we perform another set of flux calculations using the \texttt{Teukolsky} Mathematica package \cite{BHPT_TEUK}, which is also provided in the Black Hole Perturbation Toolkit.\footnote{Note that this package was designed and implemented independently of the circular flux data published in the ``Circular Orbit Self-force Data'' Toolkit data repository.} In the Mathematica code, we use anywhere from 50 to 400 digits of precision to guarantee the accuracy of the computed flux values. The fractional error between our flux interpolant and the high-precision Teukolsky fluxes are shown in the right plot of Figure \ref{fig:fluxComparison}, demonstrating strong agreement between our data and the Toolkit-generated fluxes. Therefore, we suspect that the flux values published in the ``Circular Orbit Self-force Data'' repository are not accurate to all reported digits for $ 2M \lesssim r_0 \lesssim 4 M$.

Altogether, these tests indicate that the error in our interpolated flux function is $< 10^{8}$ and often matches the intrinsic error in the underlying flux data, which was calculated to a requested precision of $\sim 10$ digits. Since flux errors scale as $\sim \epsilon^{-1}$ over an inspiral, we expect that this level of flux interpolation error will have a subradian impact on the phase accuracy of our gravitational waveforms for astrophysically-realistic systems.

\begin{figure}[!t]
    \centering
    \includegraphics[width=0.95\linewidth]{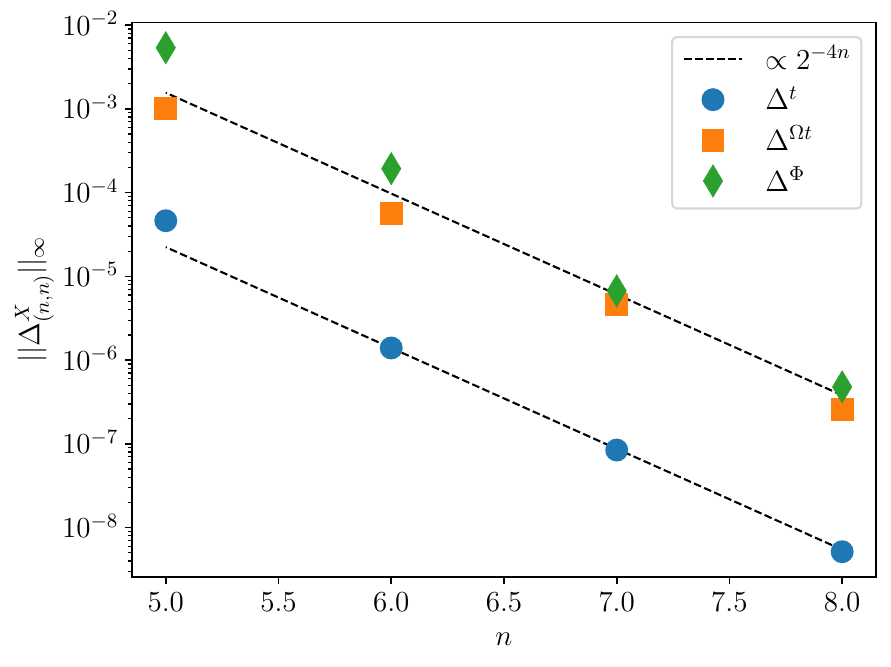}
    \caption{Convergence of the maximum absolute interpolation error for the frequency-weighted time and phase splines, $||\Delta^{\Omega t}_{(n,n)}||_\infty$ (orange squares) and $||\Delta^\Phi_{(n,n)}||_\infty$ (green diamonds), along with the fractional interpolation error for the unweighted time-spline $||\Delta^t_{(n,n)}||_\infty$ (blue diamonds). The dashed lines correspond to a power-law convergence of $2^{-4n}$.}
    \label{fig:trajConvergence}
\end{figure}

\begin{figure}[!t]
    \centering
    \includegraphics[width=0.95\linewidth]{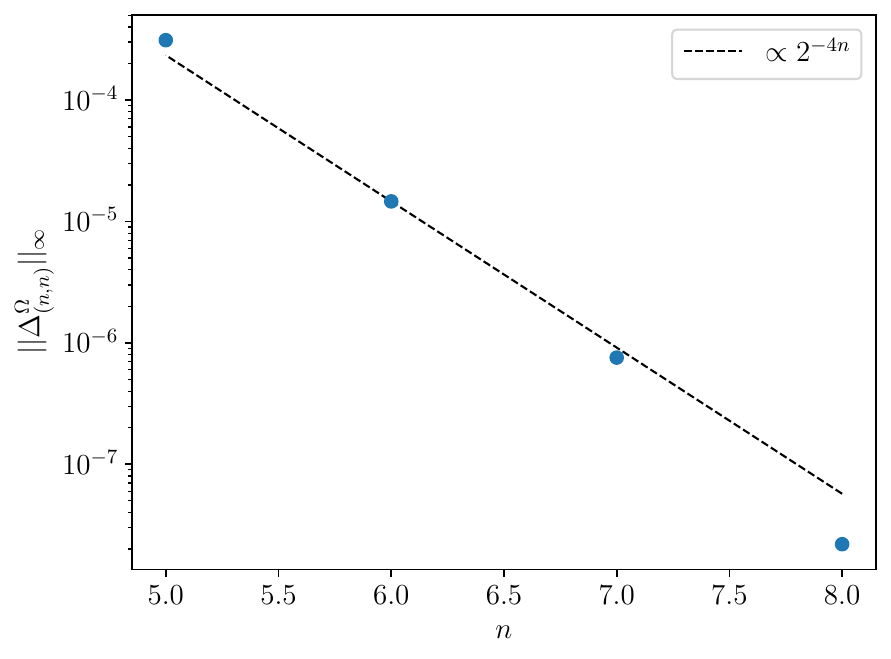}
    \caption{Convergence of the maximum fractional interpolation error for the frequency spline, $||\Delta^{\Omega}_{(n,n)}||_\infty$.}
    \label{fig:frequencyConvergence}
\end{figure}

\subsection{Numerical accuracy of trajectories}
\label{app:trajtests}

Next we validate the accuracy of our trajectories. For the evolution of $\Omega t$ and $\Phi$, we are concerned with the absolute error in the splines, since these quantities  only appear directly in the phasing of our frequency- and time-domain waveforms, respectively. However, we are concerned with the precision of $t$ when evaluating the initial phase of our waveform based on an initial frequency, i.e., $\Phi_\mathrm{initial} = \Phi(t(\Omega_0))$.

To verify the numerical convergence of our splines, we define the convergence measures,
\begin{align}
    \Delta^t_{(N',M')} &= \left| 1 - \frac{1-\check{t}^I_{(N',M')}}{1-\check{t}^I_{(N'+1,M'+1)}} \right|,
    \\
    \Delta^{\Omega t}_{(N',M')} &= \hat{\Omega}\left| {\check{t}^I_{(N',M')}} - {\check{t}^I_{(N'+1,M'+1)}} \right|,
    \\
    \Delta^\Phi_{(N',M')} &= \left| {\check{\Phi}^I_{(N',M')}} - {\check{\Phi}^I_{(N'+1,M'+1)}} \right|,
\end{align}
and plot $|| \Delta^{t,\Omega t,\Phi}_{(n,n)}||_\infty$ for $n = (5, 6, 7, 8)$ in Figure \ref{fig:trajConvergence}. As before, the error improves as we increase the sampling density by a factor of $2$ in each dimension, though in this case the maximum absolute error converges slightly faster than $2^{4n}$. Thus extrapolating this rate of convergence, we estimate $|| \Delta^{t,\Omega t,\Phi}_{(9,9)}||_\infty \lesssim 3\times 10^{-8}$.

Next, we check that the interpolated trajectories satisfy the equations of motion \eqref{eqn:eom2} via the two tests,
\begin{align}
    \delta_{t}(\hat{\Omega}, \hat{a}) &= \left| 1 - \left[1 + \frac{1}{\mathcal{F}^I_N}\left(\frac{\partial\mathcal{E}}{\partial\alpha} \right)\right]\left[1 - \frac{d\check{t}^I}{d\alpha}\right]^{-1}\right|,
    \\
    \delta_{\Phi}(\hat{\Omega}, \hat{a}) &= \left| 1 -  \left[1 + \frac{\hat{\Omega}}{\mathcal{F}^I_N}\left(\frac{\partial\mathcal{E}}{\partial\alpha} \right)\right]\left[1 -\frac{d\check{\Phi}^I}{d{\alpha}} \right]^{-1}\right|,
\end{align}
where, on the righthand side, all functions and derivatives are evaluated at $\alpha(\hat{a},\hat{\Omega})$ and $\chi(\hat{a})$. We shift the numerators and denominators by a factor $1$, because the $\alpha$-derivatives vanish at $\alpha = 0$. This translation effectively leads to $\delta_t$ and $\delta_\Phi$ measuring absolute error for values of $\alpha \lesssim 0.05$, while measuring relative error for values $\alpha \gtrsim 0.05$. Based on this analysis, we find $||\delta_t||_\infty \simeq 10^{-6.7}$ and $||\delta_\Phi||_\infty \simeq 10^{-6.8}$ when maximizing over all values of $\hat{a}$ and $\hat{\Omega}$.

To assess the precision of $\hat{\Omega}^I(\check{t};\hat{a})$, we calculate the fractional error
\begin{align}
    \Delta^\Omega_{(N',M')} &= \left| 1 - \frac{\hat{\Omega}^I_{(N',M')}}{\hat{\Omega}^I_{(N'+1,M'+1)}} \right|,
\end{align}
and plot the maximum values $||\Delta^\Omega_{(n,n)}||_\infty$ for $n = (5, 6, 7, 8)$ in Figure \ref{fig:frequencyConvergence}. With the convergence slightly better than $2^{4n}$, we estimate that $||\Delta^\Omega_{(9,9)}||_\infty < 2\times 10^{-9}$. Furthermore, we can apply the self-consistency check,
\begin{align}
    \delta_{\Omega}(\hat{\Omega}_0, \hat{a}) = \left| 1- \frac{\hat{\Omega}_0}{\hat{\Omega}^I[\check{t}^I(\hat{\Omega}_0)]} \right|,
\end{align}
and we find that the maximum error across all values is $||\delta_{\Omega}||_\infty = 3\times 10^{-9}$.

\subsection{Numerical precision and accuracy of waveform amplitudes}
\label{app:ampAccuracy}

We look at the numerical precision of the spline for the magnitude of our complex waveform amplitude, $A^I_{\ell m}$, and the accuracy of the amplitude phase spline $\psi^I_{\ell m}$. Since the grids that we interpolate are already sparsely populated we do not follow the same analysis as the previous subsections. Rather than comparing splines from successive iterations of downsampled data, we instead compute a new set of complex amplitude values on a $10 \times 30$ grid in $(\chi, \alpha)$. We perform this calculation for $(\ell,m) = [(2,2), (5,2), (5,5), (20,2), (20,20)]$ to get a representative sample of small and large $\ell$-values and small and large $m$ values. As a result, the maximum fractional errors in $A^I_{\ell m}$ are $2.1\times10^{-6}$, $6.2\times10^{-6}$, $5.2\times10^{-6}$, $5.5\times10^{-3}$, and $2.1\times10^{-5}$ for the $(2,2)$, $(5,2)$, $(5,5)$, $(20,2)$, and $(20,20)$ modes, respectively. For the maximum absolute errors in $\psi^I_{\ell m}$ we get $2.3\times10^{-6}$, $2.4\times10^{-6}$, $4.8\times10^{-6}$, $2.3\times10^{-3}$, and $1.9\times10^{-5}$ for the $(2,2)$, $(5,2)$, $(5,5)$, $(20,2)$, and $(20,20)$ modes, respectively. Therefore, $\psi_{\ell m}$ achieves subradian phase accuracy while $A^I_{\ell m}$ maintains precision to at least 3 digits even for the least dominant mode $(20, 2)$, which will only be included in waveforms with very low mode selection thresholds $\epsilon_\mathrm{mode} < 10^{-5}$.

\begin{figure*}[!htp]
    \centering
    \includegraphics[width=0.98\linewidth]{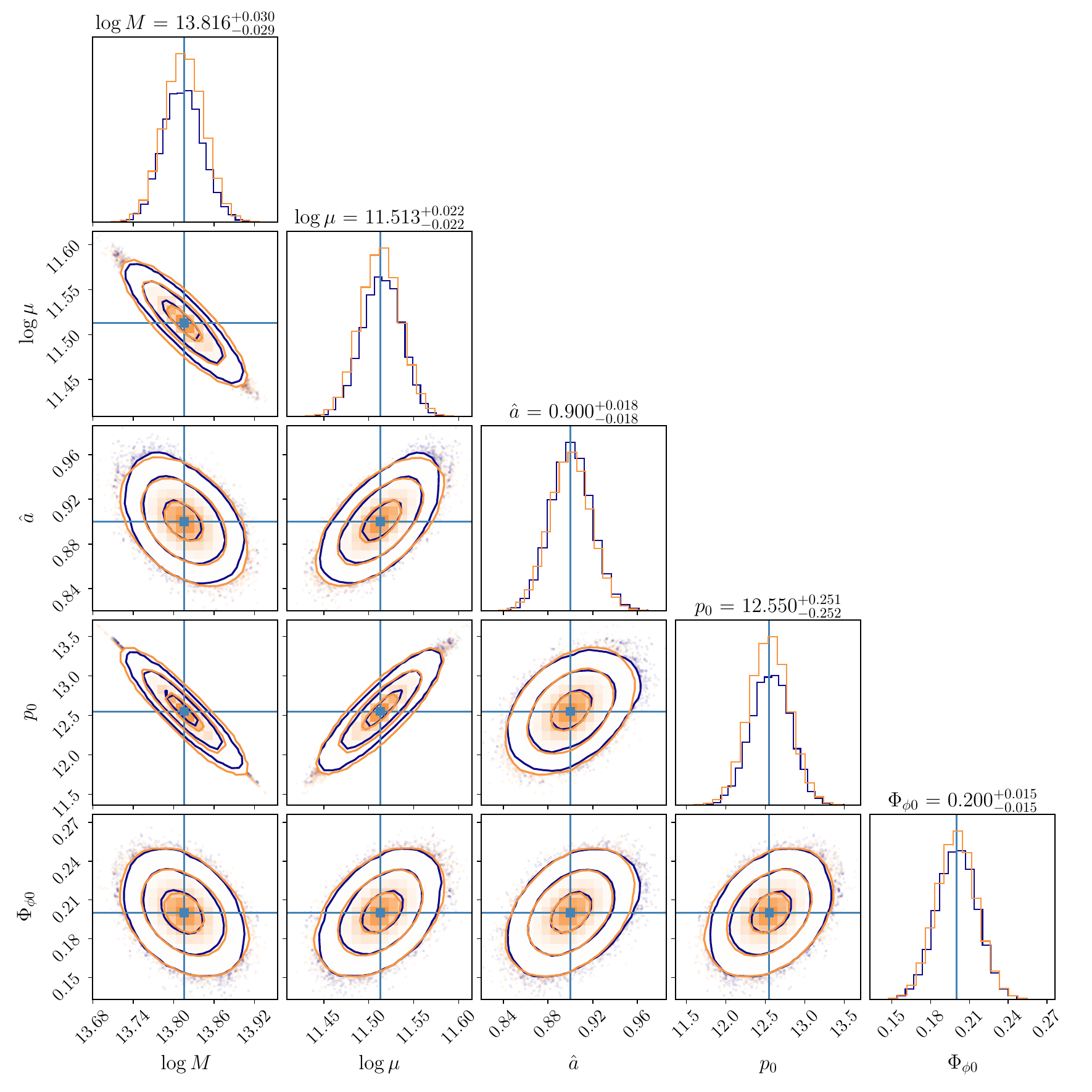}
    \caption{Posterior distribution based on an parallel-tempered MCMC sampling (dark blue lines) and a Fisher matrix analysis (lighter orange lines).}
    \label{fig:mcmcOverlay}
\end{figure*}

\section{Transformations for retrograde orbits}
\label{app:retrograde}

Consider an orbit with positive spin $\hat{a} > 0$, but negative angular momentum and orbital frequency $\hat{\Omega} < 0$ and $\mathcal{L}_z < 0$ (which is parametrized by the choice $x<0$). Under the transformation $(\hat{a}, \hat{\Omega}) \rightarrow (-\hat{a}, -\hat{\Omega})$ on the waveform, we must first understand the impact on the mode amplitudes $A_{\ell m}(\hat{a},\hat{\Omega})e^{i\psi_{\ell m}(\hat{a},\hat{\Omega})} = A_{\ell m}(-\hat{a},-\hat{\Omega}) e^{-i\psi_{\ell m}(-\hat{a},-\hat{\Omega})}$ and $\Phi(\hat{a},\hat{\Omega}) = -\Phi(\hat{a},\hat{\Omega})$. Consequently, the waveform takes the form
\begin{subequations}
    \begin{align}
    h_+(-\hat{a},-\hat{\Omega}) &= \sum_{\ell m} \mathcal{A}_{\ell m}(\hat{a},\hat{\Omega}) Y^+_{\ell m} \cos(\psi_{\ell m} - m[\phi + \Phi]),
    \\
    h_\times(-\hat{a},-\hat{\Omega}) &= \sum_{\ell m} \mathcal{A}_{\ell m}(\hat{a},\hat{\Omega}) Y^\times_{\ell m} \sin(\psi_{\ell m} - m[\phi + \Phi]),
\end{align}
\end{subequations}
or $h_{+/\times}(-\hat{a},-\hat{\Omega}; \theta, \phi) = h_{+/\times}(\hat{a},\hat{\Omega}, \pi -\theta, -\phi)$.

\section{Numerical methods for Fisher calculations}
\label{app:fisher}

To construct the Fisher information matrix in \eqref{eqn:fisher} we first compute derivatives of our waveforms in the time-domain. We take derivatives with respect to the intrinsic parameters (e.g., $\partial h_m /\partial\log M$) on a mode by mode basis via \eqref{eqn:timeDomainReduced}. Note that we take derivatives with respect to the logs of the masses, $\log M$ and $\log \mu$, rather than the masses themselves. We use a basic finite central difference stencil
\begin{align}
    \partial_i f(\theta) \approx \frac{f(\theta + h) - f(\theta - h)}{2h},
\end{align}
to differentiate $\mathcal{A}_{\ell m}$ and $\Phi_{\ell m} = \psi_{\ell m} + m(\phi - \Phi)$ at each time step. We adapt the step-size $h$ until we converge to a stable numerical approximation of the derivative. However, very small step-sizes can accentuate numerical noise inherent in our waveforms. To mitigate this noise, we apply a Savitzky–Golay filter to suppress this noise in our numerical derivatives.

Then the waveform derivative is assembled in the source frame via
\begin{subequations} \label{eqn:timeDomainDerivatives}
    \begin{align}
    \partial_i h_+ &= +\sum_{\ell m} \left[ \partial_i \mathcal{A}_{\ell m} \cos\Phi_{\ell m} - \mathcal{A}_{\ell m} \partial_i \Phi_{\ell m} \sin\Phi_{\ell m}\right] Y^+_{\ell m},
    \\
    \partial_i h_\times &= -\sum_{\ell m} \left[ \partial_i \mathcal{A}_{\ell m} \sin\Phi_{\ell m} + \mathcal{A}_{\ell m} \partial_i \Phi_{\ell m} \cos\Phi_{\ell m}\right] Y^\times_{\ell m},
    \end{align}
\end{subequations}
and then we transform to the SSB frame. For derivatives with respect to $\Phi_{\phi 0}$, rather than using numerical derivatives, we use the analytic solutions $\partial_{\Phi_{\phi 0}} C_{\ell m} = 0$ and $\partial_{\Phi_{\phi 0}} \Phi_{\ell m} = -m$.

For derivatives with respect to the extrinsic parameters, we use a higher-order numerical finite central difference stencil,
\begin{align}
    \partial_i f(\theta) &\approx -\frac{1}{12h}f(\theta + 2h) + \frac{2}{3h}f(\theta + h)
    \\ \notag
    & \qquad \qquad \qquad - \frac{2}{3h}f(\theta - h) + \frac{1}{12h}f(\theta-2h),
\end{align}
which we apply to the full time-domain waveform in the SSB waveform. For derivatives with respect to $d_L$, we replace numerical derivatives with the analytic solution $\partial_{d_L} h_\mathrm{SSB} = -h_\mathrm{SSB}/d_L$. Finally, to construct $\Gamma^{ij}$, we apply a Tukey window to the numerical derivatives prior to applying the DFT. We then throw away the last 50 frequency bins of the Fourier transform to remove any residual high-frequency noise. We find that this is essential for improving the numerical stability of our Fisher matrix analysis. We then evaluate the inner products to determine the components of $\Gamma^{ij}$.

Due to high condition numbers of EMRI Fisher matrices, $\Gamma^{ij}$ is very nearly a singular matrix, and thus inverting the Fisher matrix can also be a numerically unstable process. Thus, we instead construct the \emph{pseudoinverse} $[\tilde{\Gamma}^{-1}]^{ij}$, which is well defined for singular matrices and satisfies the relation
\begin{align} \label{eqn:pinvRelation}
    \Gamma^{ai}[\tilde{\Gamma}^{-1}]^{ij} \Gamma^{jb} = \Gamma^{ab}.
\end{align}
For nonsingular matrices $[\tilde{\Gamma}^{-1}]^{ij} = [{\Gamma}^{-1}]^{ij}$. We numerically compute $[\tilde{\Gamma}^{-1}]$ using a singular value decomposition (SVD),
\begin{align}
    \Gamma = U S V^T,
\end{align}
where $U$ and $V$ are unitary matrices and $S$ is a diagonal matrix of the form $S = \mathrm{diag}(s_0, s_1, \cdots, s_{N-1})$ where some of the values $s_i$ may be zero, which would indicate that $\Gamma$ is singular. The pseudoinverse is then given by
\begin{align}
    \tilde{\Gamma}^{-1} = V \tilde{S}^{-1} U^T,
\end{align}
where $\tilde{S}^{-1} = \mathrm{diag}(\tilde{s}_0^{-1}, \tilde{s}_1^{-1}, \cdots, \tilde{s}_{N-1}^{-1})$ is the pseudoinverse of $S$. The elements are given by
\begin{subequations}
    \begin{align}
    \tilde{s}_i^{-1} &= s_i^{-1}, & |s_i| &> 0,
    \\ \label{eqn:singularInversion}
    \tilde{s}_i^{-1} &= 0, & |s_i| &= 0.
\end{align}
\end{subequations}
To numerically compute $\tilde{S}^{-1}$ we instead replace \eqref{eqn:singularInversion} with $\tilde{s}_i^{-1} = 0$ for $|s_i| < \epsilon_\mathrm{SVD}$, for some numerical tolerance $\epsilon_\mathrm{SVD}$. We then vary $\epsilon_\mathrm{SVD}$ until we find a numerical solution $\tilde{\Gamma}^{-1}_{\epsilon_\mathrm{SVD}}$ that best satisfies \eqref{eqn:pinvRelation}.

To check these calculations, we verify that $\tilde{\Gamma}^{-1}$ presents a good representation of the covariances between the model parameters. In the neighborhood of $\theta^i_\mathrm{peak}$ (where the likelihood peaks), the posterior distribution $p(\theta | d=h_m(\theta_\mathrm{peak}))$ is described by a multivariate normal distribution
\begin{align}
    &-\frac{1}{2}(h_m(\theta_\mathrm{peak})-h_m(\theta)| h_m(\theta_\mathrm{peak})-h_m(\theta)) =
    \\ \notag
    &\qquad \qquad \qquad \qquad -\frac{1}{2}(\theta^{i} - \theta_\mathrm{peak}^i)[\Sigma^{-1}]^{ij}(\theta^{j} - \theta_\mathrm{peak}^j),
\end{align}
where $\Sigma \approx \tilde{\Gamma}^{-1}$ are the covariances between the model parameters. To verify the relation $\Sigma \approx \tilde{\Gamma}^{-1}$, we look to perturb two model parameters $\theta^A = \theta^A_\mathrm{peak} + \Delta \theta^A$ and $\theta^B= \theta^B_\mathrm{peak} + \Delta \theta^B$ so that we move one-$\sigma$ away from the peak of our posterior distribution. This amounts to finding values $\Delta\theta^A$ and $\Delta\theta^B$ that satisfy
\begin{align}
    2\Delta \theta^{A}\Delta\theta^{B} \tilde{\Gamma}^{AB} + \Delta \theta^{A}\Delta\theta^{A} \tilde{\Gamma}^{AA} + \Delta \theta^{B}\Delta\theta^{B} \tilde{\Gamma}^{BB} = 1,
\end{align}
where $\tilde{\Gamma} = \Gamma$ is the pseudoinverse of $\tilde{\Gamma}^{-1}$. Picking $\Delta \theta^{A} = \left[\tilde{\Gamma}^{AA}\right]^{-1/2}$ then uniquely determines $\Delta \theta^{B} = -2 \Delta\theta^{A}  \tilde{\Gamma}^{AB}/\tilde{\Gamma}^{BB}$. We then verify that for all values of $A$ and $B$ (e.g., $A = \log M$ and $B = a$)
\begin{align} \label{eqn:covCheck}
    (h_m(\theta_\mathrm{peak})-h_m(\theta_{(AB)})| h_m(\theta_\mathrm{peak})-h_m(\theta_{(AB)})) \approx 1,
\end{align}
where $\theta^i_{(AB)}$ represents the set of parameters that have been perturbed away from $\theta^i_\mathrm{peak}$ by $\Delta \theta^A$ and $\Delta \theta^B$. For all of the Fisher matrix calculations performed in this work, we find \eqref{eqn:covCheck} is satisfied to a precision $< 10^{-1}$ across all combinations of model parameters and to a precision $<10^{-2}$ for all intrinsic parameters.

As a final check, we perform a parallel-tempered Markov Chain Monte Carlo (MCMC) sampling of the posterior distribution for intrinsic the source parameters $(M, \mu, \hat{a}, p_0, \Phi_{\phi0}) = (10^6 M_\odot, 10^5 M_\odot, 0.9, 12.55, 0.2)$ using the open-source \texttt{Eryn} sampling tool \cite{ForeETC13, KarnETC23,KatzKarnKors23}.\footnote{Note that we choose a very comparable mass ratio to speed-up waveform evaluation and the calculation of the likelihood. This allows us to perform the sampling within a day on a laptop. More realistic EMRI systems are much more computationally expensive to sample and beyond the scope of this work. MCMC sampling of astrophysically-realistic EMRIs with \texttt{bhpwave} will be presented in a following paper.} To simplify the calculation, we only sample over the intrinsic parameters $(\log M, \log \mu, a, p_0, \Phi_{\phi0})$ and hold all other parameters fixed. We then perform our Fisher analysis on the same system. In Figure \ref{fig:mcmcOverlay} we plot the two dimensional contours of the posterior distribution calculated from our MCMC sampling (dark blue lines). Each contour line corresponds to a one-$\sigma$ deviation from a neighboring contour. We then use the covariances predicted from our Fisher analysis to sample a multivariate distribution and overlay this on top of our MCMC results (lighter orange lines). As we can see, the two sets of samples lie nearly on top of one another, indicating that our Fisher matrix analysis provides a leading-order estimate for the shape of the posterior.

\bibliography{parent}

\end{document}